\documentclass[preprint]{aastex631}

\let\tablenum\relax
\usepackage{stix, isomath, siunitx}
    \DeclareSIUnit\bar{bar}
    \DeclareSIUnit\angstrom{\text{Å}}
\usepackage[version=4]{mhchem}

\shorttitle{Europa/Ganymede/Callisto Aurora}
\shortauthors{de~Kleer et al.}
\graphicspath{{./}{figures/}}

\begin{document}

\title{The Optical Aurorae of Europa, Ganymede and Callisto}

\author[0000-0002-9068-3428]{Katherine de Kleer}
\affiliation{California Institute of Technology, 1200 E California Blvd, MC 150-21, Pasadena CA 91125, USA}

\author[0000-0001-5683-0095]{Zachariah Milby}
\affiliation{California Institute of Technology, 1200 E California Blvd, MC 150-21, Pasadena CA 91125, USA}

\author[0000-0002-6917-3458]{Carl Schmidt}
\affiliation{Boston University, 725 Commonwealth Ave, Boston MA 02215, USA}

\author[0000-0003-3887-4080]{Maria Camarca}
\affiliation{California Institute of Technology, 1200 E California Blvd, MC 150-21, Pasadena CA 91125, USA}

\author[0000-0002-8255-0545]{Michael E. Brown}
\affiliation{California Institute of Technology, 1200 E California Blvd, MC 150-21, Pasadena CA 91125, USA}

\begin{abstract}

The tenuous atmospheres of the Galilean satellites are sourced from their surfaces and produced by a combination of plasma-surface interactions and thermal processes. Though thin, these atmospheres can be studied via their auroral emissions, and most work to date has focused on their aurora at UV wavelengths. Here we present the first detections of Ganymede's and Callisto's optical aurorae, as well detections of new optical auroral lines at Europa, based on observations of the targets over ten Jupiter eclipses from 1998 to 2021 with Keck/HIRES. We present measurements of \ce{O}\,I emission at \SIlist[parse-numbers=false]{6300/6364;5577;7774;8446}{\angstrom} and place upper limits on hydrogen at \SI{6563}{\angstrom}. These constitute the first detections of emissions at \SIlist{7774;8446}{\angstrom} at a planetary body other than Earth. The simultaneous measurement of multiple emission lines provides robust constraints on atmospheric composition. We find that the eclipse atmospheres of Europa and Ganymede are composed predominantly of \ce{O2} with average column densities of \SI[parse-numbers=false]{(4.1 \pm 0.1)\times 10^{14}}{cm^{-2}} and \SI[parse-numbers=false]{(4.7 \pm 0.1)\times 10^{14}}{cm^{-2}}, respectively. We find weak evidence for \ce{H2O} in Europa's bulk atmosphere at an $\ce{H2O}/\ce{O2}$ ratio of $\sim$0.25, and place only an upper limit on \ce{H2O} in Ganymede's bulk atmosphere, corresponding to $\ce{H2O}/\ce{O2} < 0.6$. The column density of \ce{O2} derived for Callisto is \SI[parse-numbers=false]{(4.0 \pm 0.9)\times 10^{15}}{cm^{-2}} for an assumed electron density of \SI{0.15}{cm^{-3}}, but electron properties at Callisto's orbit are very poorly constrained.

\end{abstract}

\section{Introduction} \label{sec:intro}

The tenuous atmospheres of the Galilean satellites Europa, Ganymede, and Callisto are composed of a combination of \ce{O2}, \ce{O}, \ce{H2O}, and \ce{CO2}, and are ultimately sourced from their surfaces. The detailed compositions of these atmospheres, along with their spatial distributions and temporal variabilities, thus provide information on their surface compositions and on surface modification processes such as sputtering. 

Europa's atmosphere was first detected via the \SIlist{1356;1304}{\angstrom} multiplets of atomic oxygen in emission using the Goddard High Resolution Spectrograph (GHRS) on the Hubble Space Telescope (HST) \citep{hall1995}; additional measurements of Europa's aurora, and a first detection of the same auroral lines at Ganymede, were subsequently made \citep{hall1998}. The emission was attributed to electron impact dissociative excitation of predominantly \ce{O2} for both Europa and Ganymede, with perhaps some contribution from atomic \ce{O} especially in the case of Ganymede, on the basis of the ratio of emission between the two multiplets. The strength of the emission lines corresponds to an \ce{O2} atmosphere of \SIrange{2.4e14}{14e14}{cm^{-2}} at Europa and \SIrange{1e14}{10e14}{cm^{-2}} at Ganymede, using \textit{Voyager}- and \textit{Galileo}-based estimates for the density and temperature and the exciting electrons \citep{hall1995,hall1998}. 

Various mechanisms have been proposed for the production of these tenuous atmospheres. Motivated by an early claim of a \SI{1}{\micro\bar} atmosphere at Ganymede \citep{carlson1973}, \cite{yung1977} proposed a sublimation-driven water atmosphere that produces a molecular \ce{O2} atmosphere due to photolysis of \ce{H2O} followed by the escape of the lighter hydrogen. However, this model predicts a surface density well above the subsequent \textit{Voyager 1} findings \citep{broadfoot1979,wolff1983}. \cite{lanzerotti1978} proposed that Ganymede's atmosphere is produced by jovian plasma bombardment, which sputters neutral molecules off the icy surface, and argued that this is a more efficient mechanism for producing \ce{O2} especially at the low end of plausible surface temperatures. \cite{brown1980} then measured ion sputtering yields as a function of ice temperature, and more recent modeling work has calculated sputtering yields of O$_2$ and H$_2$ for Europa \citep[][]{cassidy2013} and for icy satellites in general \citep[][]{teolis2017}. Thermal \ce{H2} and \ce{O2} products do not condense as efficiently as sputtered water molecules, and \ce{H2} readily escapes, leaving \ce{O2} as the primary constituent.  \cite{johnson1981,johnson1982} used these experiments to predict that a $\sim10^{14}$ cm$^{-2}$ atmosphere of \ce{O2} would be sustained at Ganymede and $\sim10^{15}$ cm$^{-2}$ at Europa, and also calculated an expected $\sim$\SIrange{2e12}{1e14}{cm^{-2}} of \ce{H2O} across the three icy satellites. However, models subsequently showed that \ce{O2} yields from ion sputtering are insufficient to match the observed column densities, unless additional ``resputtering'' by freshly ionized \ce{O2} augments the flux from Jupiter's magnetosphere \citep{ip1996}. \cite{saur1998} modeled the atmospheric sources and sinks, including plasma deflection, and found that a combination of supra-thermal torus ions and thermal ions sputtering \ce{O2} from the surface could produce a stable atmosphere around the measured column densities, and demonstrated that the re-sputtering mechanism could not contribute a significant amount.

Images of Europa and Ganymede in the UV \ce{O}\,I multiplets (\SIlist{1304; 1356}{\angstrom}) with HST/STIS have yielded evidence for spatial variations across both satellites \citep{mcgrath2013,feldman2000,roth2016}. For Europa the spatial distribution of emissions exhibits some systematic trends but is still not fully understood. A dusk-dawn asymmetry is observed in both the UV and optical auroral data \citep{roth2016,dKB2019}, which is consistent with simulations \citep{oza2019} and suggests a thermal role in the production of the \ce{O2} atmosphere \citep{oza2018,johnson2019}. For Ganymede, the aurora appear to behave analogously to Earth's auroral ovals, whereby electrons are accelerated into the near-surface region along field lines at the open-closed field line boundary \citep{mcgrath2013,feldman2000,eviatar2001_aurora}. The oscillation amplitude of the ovals has been used as evidence to support the existence of a subsurface ocean on Ganymede \citep{saur2015}. The brightness of Ganymede's auroral spots cannot be matched by models assuming the electron energies and densities at Ganymede's orbit, requiring either higher electron energies or higher-density, lower-temperature electrons \citep{eviatar2001_aurora}; the higher electron energies are attributed to local acceleration at Ganymede \citep{eviatar2001_aurora}. The uncertainties on the electron population exciting the emissions permit \ce{O2} column densities in the range of \SIrange{1e14}{30e14}{cm^{-2}} for Ganymede. In contrast, Europa's auroral emissions are consistent with excitation by thermal magnetospheric electrons and do not require local acceleration \citep{saur1998}.

Recently, \cite{roth2021_europa} and \cite{roth2021_ganymede} used auroral data sets to independently constrain the atomic \ce{O} abundance in the atmospheres of Europa and Ganymede by measuring the resonant scattering component of the \SI{1304}{\angstrom} emission as the satellites passed through Jupiter's shadow. These data sets put a tight upper limit on the \ce{O} abundance, which then requires a new mechanism to explain the low \SI[parse-numbers=false]{1356/1304}{\angstrom} ratio (sometimes referred to as $r_{\gamma}$(OI), e.g. \citep{roth2021_europa}) on these satellites' trailing hemispheres. The proposed mechanism is a consistently-present \ce{H2O} atmosphere centered on the trailing hemisphere for Europa and both hemispheres (though $6\times$ denser on the trailing) for Ganymede, which can be produced by sublimation in the case of Ganymede \citep[as is also predicted on Callisto;][]{carberry-mogan2021} and by sputtering combined with sublimation of the fresh deposits of sputtered \ce{H2O} in the case of Europa \citep{roth2021_europa,teolis2017}. The derived mixing ratios of $\ce{H2O}/\ce{O2}$ over the trailing hemisphere are in the 10 to 30 range for both satellites. The UV lines are significantly more sensitive to \ce{O2} than \ce{H2O}, and observations of \ce{H} in addition to \ce{O}, and/or observations of lines with higher intrinsic emission rates following electron impact on \ce{H2O}, would strengthen constraints on \ce{H2O} presence and abundance. 

To date, studies of the auroral emissions of Europa and Ganymede have been conducted almost exclusively in the UV. However, measurements of additional lines at different wavelengths can provide more robust constraints on atmospheric composition, and have the potential to reveal additional processes at work or even new atmospheric constituents. The optical auroral lines have been studied at Io for decades \citep[e.g.][]{belton1996,schmidt2022}. On the icy Galilean satellites, whose atmospheres and orbital environments are both less populated than those of Io, the first published detection of the optical aurora was the \SIlist[parse-numbers=false]{6300/6364}{\angstrom} oxygen emission observed at Europa from HST/STIS and the Keck High Resolution Spectrograph (HIRES) \citep{dKB2018,dKB2019}. The data supported an \ce{O2} atmospheric composition and displayed a high level of temporal variability and spatial patchiness on top of an overall asymmetry in the auroral brightness with more emission on the trailing/dusk side, consistent with the UV morphology observed by \cite{roth2016}.

Here we present observations of Europa, Ganymede, and Callisto in Jupiter eclipse, over ten eclipses total between 1998 and 2021, taken with Keck/HIRES. The data cover wavelengths from \SIrange{5000}{9000}{\angstrom}. We present measurements of the \SIlist[parse-numbers=false]{6300/6364;5577;7774;8446}{\angstrom} \ce{O}\,I emissions from these satellites. All emission lines listed are detected on at least one occasion, and we present upper limits on dates when a given line is not detected, as well as on H$\mathrm{\alpha}$ \SI{6563}{\angstrom} on all dates. The observations and data reduction procedures are described in Section \ref{sec:obs}. The model used to interpret the data is expanded from that of \cite{dKB2018} and is described in Section \ref{sec:analysis}, including the MCMC retrieval algorithm. The derived auroral brightnesses are presented and discussed in Section \ref{sec:results}, and conclusions are summarized in Section \ref{sec:conc}.

\section{Observations and Data Reduction} \label{sec:obs}

\begin{deluxetable*}{lllcccccl}
\tablecaption{Observing parameters.\label{tbl:obs}}
\centering
\tablehead{
Date [UTC] & Time [UTC] & Target & $t_\text{int}$ [min] & Diam. [arcsec] & CML\textsuperscript{a} [\si{\degree}{W}] & BG\textsuperscript{b} & Avg. Sep.\textsuperscript{c} & Notes
}
\startdata
1998-11-15 & 05:56--08:55 & Ganymede & 40 & 1.63 & 10--16 & 1.5 & 2.2\\
2018-03-22\textsuperscript{d} & 12:27--14:15 & Europa & 50 & 0.91 & 350--357 & 38 & 0.7 & Poor weather \\
2018-06-15 & 07:10--08:54 & Ganymede & 20 & 1.58 & 7--11 & 18 & 1.0\\
2021-05-20 & 13:40--15:05 & Europa & 40 & 0.87 & 345--351 & 5.4 & 1.2\\
2021-06-08 & 12:48--15:17 & Ganymede & 85\textsuperscript{e} & 1.55 & 348--353 & 2.1 & 2.1\\
2021-06-21 & 13:05--15:15 & Europa & 60 & 0.96 & 345--354 & 5.7 & 0.9\\
2021-07-04 & 14:13--15:25 & Callisto & 25 & 1.54 & 349--350 & 1 & 3.7\\
2021-07-16 & 10:12--11:39 & Europa & 50 & 1.03 & 348--354 & 14 & 0.6\\
2021-09-26 & 09:05--11:03 & Callisto & 20 & 1.58 & 5--9 & 1.7 & 2.4 & Poor weather \\
2021-10-01 & 04:55--08:19 & Ganymede & 70 & 1.70 & 7--14 & 2.3 & 1.3 & Poor weather
\enddata
\vspace*{3pt}
\textsuperscript{a}{Central meridian longitude, or sub-observer longitude.}\\
\textsuperscript{b}{Sky background level measured near \SI{6300}{\angstrom}, relative to a measured \SI{21}{electrons.s^{-1}.arcsec^{-2}} background in the Callisto observations taken July 4, 2021.}\\
\textsuperscript{c}{Average angular distance of the target from Jupiter's limb in units of Jupiter radii.}\\
\textsuperscript{d}{Previously published in \cite{dKB2018}.}\\
\textsuperscript{e}{Some infrared orders are removed during averaging due to additional contamination; total integration times were 70 minutes for \SI{7774}{\angstrom} and 60 minutes for \SI{8446}{\angstrom}.}\\

\end{deluxetable*}

Observations of Europa, Ganymede, and Callisto in Jupiter eclipse were obtained on 4, 4, and 2 occasions, respectively, between 1998 and 2021 using the HIRES instrument \citep{vogt1994} on the Keck I telescope at the summit of Maunakea in Hawaii. Observations in eclipse remove the reflected sunlight component which would otherwise overwhelm the signal from the visible-wavelength aurora. One eclipse observation of Ganymede is from 1998, which utilized HIRES pre-upgrade when it had a single CCD and less extensive spectral coverage. Table \ref{tbl:obs} lists the observing parameters for each of these ten nights. The observing sequence typically consisted of 300-second integrations on the satellite while in eclipse, alternated with offsets to a nearby sunlit pointing satellite to check positioning and recenter if needed. The total on-target integration time used in the analysis for each observation is given in Table \ref{tbl:obs}.

The HIRES post-upgrade three CCD mosaic covers wavelengths from roughly \SIrange{5000}{9500}{\angstrom} with our observing set-up. The echelle and cross-disperser angles were chosen to ensure that the lines of interest did not fall in the gaps between orders. A slit width of \SI{1.722}{\arcsec} was used to cover the entire satellite, resulting in a spectral resolution of approximately \num{24000}. Slit lengths of \SI{7}{\arcsec} and \SI{14}{\arcsec} were alternately used; a longer slit length increases the amount of sky available for sky subtraction but results in overlap between the orders at short wavelengths.

To correct for cosmic rays, we used the Laplacian cosmic-ray identification method L.A.Cosmic \citep{vanDokkum2001} as implemented in the Astropy-affiliated package \texttt{ccdproc} using the methodology described in \cite{mccully18}. We then bias-subtracted, flat-fielded and gain-corrected all science images. We used the sharp edges in the flat-field to identify the boundaries of the individual echelle orders which we used to rectify each order. For the wavelength calibrations we used thorium-argon arc lamp lines to calculate a wavelength solution for each echelle order by fitting third-degree polynomials. We also corrected for airmass extinction using the wavelength-dependent magnitude attenuation appropriate for the summit of Maunakea derived by \cite{Buton2013}. To subtract the background, we calculated a characteristic, normalized spatial profile along the slit in the area immediately around the emission line, excluding pixels contained within an aperture covering the target satellite. We then fit this profile and a constant term using ordinary-least-squares to subtract the background from each echelle order.

To calibrate each observation from \si{electrons.s^{-1}.bin^{-1}} to rayleighs (R), we calculated an expected spectral surface brightness for Jupiter's central meridian using its absolute reflectivity \citep{Woodman1979} and a solar reference spectrum scaled to the Jupiter-Earth distance at the time of the observations. We then calculated the total number of electrons per second per arcsecond from Jupiter at each auroral wavelength, which provides a direct conversion from electrons to photons. To extract the target surface brightness, we calibrate the data using the apparent angular size of the target disk and the Jupiter flux calibration described above, then fit a Gaussian model and integrate it over a window of $\pm\SI{1}{\angstrom}$ around the line center (or each line center for a multiplet emission). We calculate brightnesses from both individual frames and an average of all frames.

\begin{figure*}
    \centering
    \includegraphics[width=\textwidth]{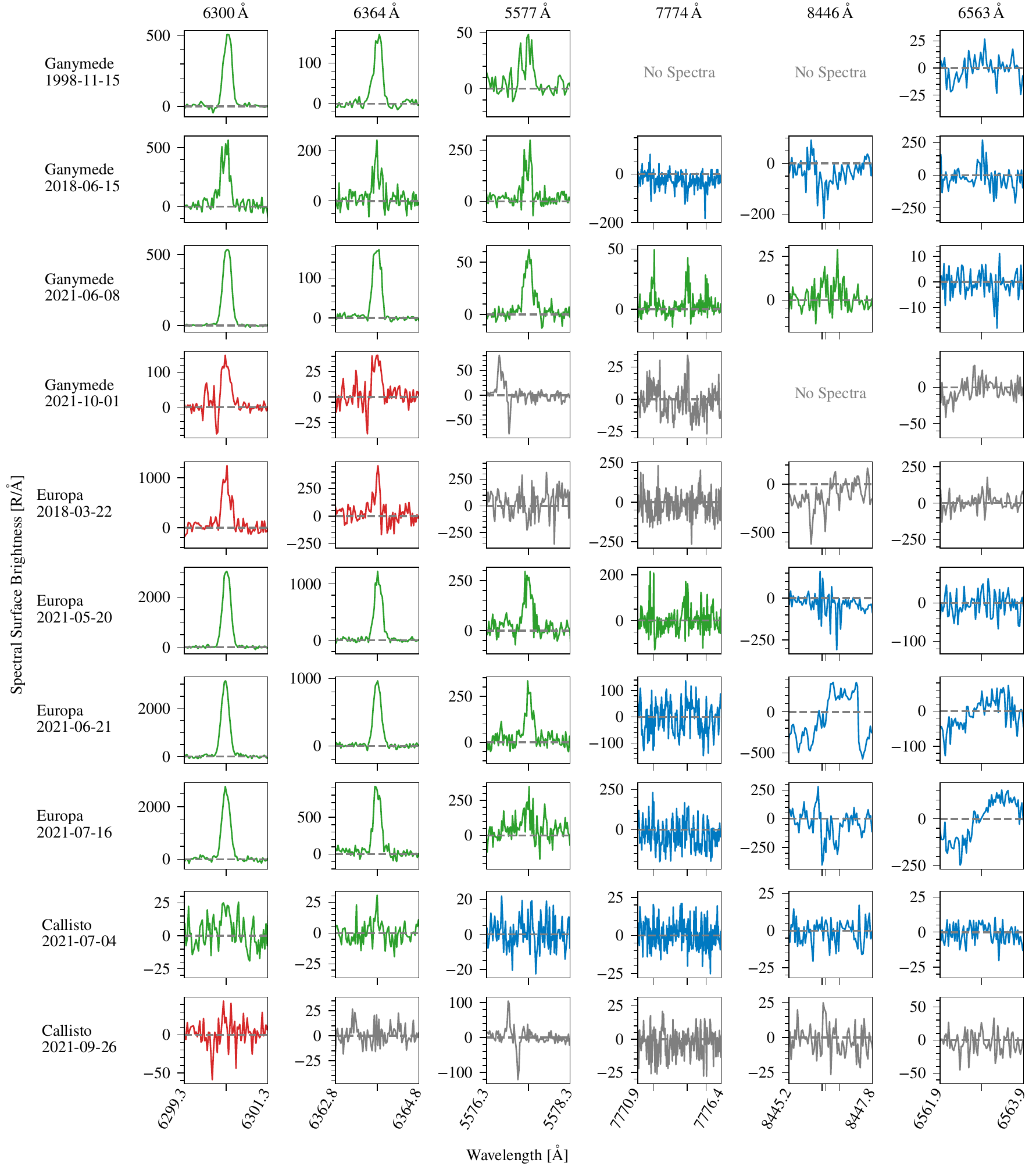}
    \caption{Average spectra centered on each emission line discussed in this paper on each night of observation. All spectra are reported in \si{R.\angstrom^{-1}} divided by the angular size of the target. Green spectra are detections with calibrated line strengths; red spectra indicate cloudy nights where a detection is made but no quantitative measurement is reported; blue spectra indicate upper limits are reported; and black spectra are non-detections where no upper limit is reported due to non-photometric conditions. Ticks along the horizontal axis shows the location of the center of the emission line(s) and the numbers on the bottom row indicate the wavelength boundaries of each plot in that column. For single emission lines, the boundaries are \SI{\pm 1}{\angstrom} from the line center. For triplet lines the boundaries are \SI{-1}{\angstrom} from the shortest wavelength and \SI[parse-numbers=false]{+1}{\angstrom} from the longest wavelength. All emission lines arise from atomic oxygen except for hydrogen at \SI{6563}{\angstrom} (which is not detected).}
    \label{fig:allspectra}
\end{figure*}

The reduced and calibrated spectra in the vicinity of each measured emission line on each date are shown in Figure \ref{fig:allspectra}.

In order for a detection to be reported instead of an upper limit, the following conditions must be met:
\begin{enumerate}
    \item The fitted surface brightness indicates a detection at the 2$\mathrm{\sigma}$-level or better,
    \item The emission line is centered at the expected Doppler-shifted wavelength given the line-of-sight motion of the target,
    \item The emission is localized along the slit, and
    \item The spectrum passes a visual inspection to ensure no false detections are reported due to poor background subtraction or other systematic contributions.
\end{enumerate}
For cloudy or non-photometric nights, we report whether or not a detection was made but do not give quantitative measurements or upper limits due to the uncertainty on the flux calibration. Figure \ref{fig:allspectra} color-codes these cases as described in the caption.

\section{Analysis}\label{sec:analysis}
\subsection{Emission Model}\label{sec:model}
All observations were made with the target satellites in Jupiter eclipse, which removes any ambiguity about the relative roles of photo-excitation vs. electron excitation. Our model assumes that all auroral emissions are produced by electron impact direct excitation or dissociative excitation of oxygen-bearing species. The model includes the parent species \ce{O}, \ce{O2}, \ce{H2O} and \ce{CO2}. The strength of the observed emission lines is a function of the satellite's atmospheric composition and density, and the density and energy distribution of the incident electrons. Our model is adapted from that described in \cite{dKB2018} and uses experimentally-derived emission cross-sections. 
The electron energies and temperatures for Europa's orbital location are taken from \cite{bagenal2015} and are derived from \textit{in situ} measurements; we use an electron density of \SI{160}{cm^{-3}} with Maxwellian-distributed energies, with 95\% of the electron population centered at \SI{20}{eV} and 5\% centered at \SI{250}{eV}. Note that older observational studies, and modeling papers that interpret these observations, used an electron density $4\times$ lower \citep[e.g.][]{hall1995,hall1998,roth2016,vorburger2021}, and column densities from these works need to be scaled down by a factor of 4 for comparison with our results and with other more recent work \citep{dKB2018,dKB2019,roth2021_europa}. The model of \cite{oza2019} uses an electron density of \SI{70}{cm^{-3}}, intermediate between the two values used in other work.

The electrons at Ganymede's orbit were similarly observed by \textit{Voyager} to be composed of two populations, with a cold population at \SIrange{30}{60}{eV} and a hot population at \SI{200}{eV} constituting 10\% of the total number density of \SIrange{3}{10}{cm^{-3}} \citep{sittler1987}. However, \cite{eviatar2001_aurora} demonstrate that the electron populations at Ganymede's orbit are insufficient to explain the brightness of the UV aurora and postulate that the aurora are excited by electrons locally accelerated to \SI{100}{eV}. We follow these and other previous authors in adopting a Maxwellian distribution of electron energies centered at \SI{100}{eV}, with an electron density of \SI{20}{cm^{-3}} \citep[though note that some modeling work has used a higher density of \SI{70}{cm^{-3}}; ][]{leblanc2017}. An electron density of \SI{20}{cm^{-3}} is consistent with recent \textit{in situ} measurements from the Ganymede \textit{Juno} fly-by, which took place one day prior to our 2021-06-08 Ganymede eclipse observation \citep{kurth2022}. Although the electron properties are uncertain and therefore the derived column densities should be viewed with caution, the ratios between emission lines do not change substantially with adopted electron properties and the relative abundances of different species is therefore robust to this uncertainty.

For Callisto, information on the electron energies and densities is even more limited. \textit{Voyager} measured an electron density of \SIrange{0.1}{1.0}{cm^{-3}} at Callisto's orbit \citep{belcher1983}, while \textit{Galileo} found \SIrange{0.01}{1.0}{cm^{-3}} depending on plasma sheet distance \citep{kivelson2004}. The electron temperature estimates from the above works range from \SIrange{35}{200}{eV}, with a supra-thermal population. We adopt a density of \SI{0.15}{cm^{-3}} at \SI{35}{eV} following \cite{belcher1983}. This is the same electron density adopted by \cite{cunningham2015}, but they included a hot population at \SI{1.5}{keV} constituting 27\% of their total population. The cross-section measurements do not extend to this energy, but we test adding a hot population at the same ratio at either \SI{250}{eV} and \SI{1.5}{keV} (extrapolating the cross-sections beyond the measurement energies), and find that derived column densities change by up to \num{\pm 30}\% depending on the electron energies. This is well within the uncertainty in the overall electron densities, so we neglect the hot electron population in our modeling. 

In addition to extending the model to Callisto, we update the model of \cite{dKB2018} by adding \ce{CO2} as a parent molecule. Emission cross-sections for electron impact dissociative excitation of \ce{CO2} are not available for several of the UV/optical \ce{O}\,I emission lines, and for \ce{CO2} our model therefore includes only emission from the \ce{O(^1S)}, \ce{O(^3S)} and \ce{O(^5S)} states at \SIlist{5577;1304;1356}{\angstrom} respectively. We adopt the emission cross sections for \ce{O(^1S)} recommended by \cite{itikawa2002}, which are based on the measurements of \cite{leclair1994}. We adopt the cross sections for \ce{O(^3S)} from \cite{ajello1971}, but note that their measurements disagree with those of \cite{mumma1972} even after revised normalization of the latter \citep{itikawa2002}, and the cross sections are consequently uncertain at the $\sim$10\% level over most electron energies and even higher at the low end of the electron energy distribution. \cite{ajello1971} also measured the emission from the \ce{O(^5S)} state at \SI{1356}{\angstrom}. Their emission cross sections are presented relative to those of \SI{1304}{\angstrom} and are therefore limited by the propagated uncertainties from the \ce{O(^3S)} measurements. Additionally, the authors note that the \SI{1356}{\angstrom} emission is a lower limit since the \ce{O(^5S)} atom may acquire excess kinetic energy during dissociation beyond the thermal velocity. The limitations on the above measurements, and the lack of measurements for other emission lines, highlights the need for future lab measurements of these cross-sections for interpretation of aurora at \ce{CO2}-containing atmospheres throughout the solar system.

Our updated auroral model now also includes the \ce{O}\,I emissions at \SIlist{7774;8446}{\angstrom} from parent molecules O, \ce{O2} and \ce{H2O}. Both emission features are triplets that are produced by the atomic oxygen transitions \ce{({3p}\,^5P)\to ({3s}\,^5S^$\circ$)} and \ce{({3p}\,^3P)\to ({3s}\,^3S^$\circ$)} respectively. Their radiative cascades to ground produce the well known \SI{1356}{\angstrom} and \SI{1304}{\angstrom} UV lines, and so the cascade contributions into these UV lines is simply equal to the \SIlist{7774;8446}{\angstrom} brightnesses in Rayleigh units. The cross-sections for both emissions following electron impact on \ce{O2} are taken from \cite{schulman1985}; on \ce{H2O} from \cite{beenakker1974} following the recommendation of \cite{itikawa2005}; and on \ce{O} from the excitation cross-sections recommended by \cite{laher1990}, which are based on the measurements of \cite{gulcicek1988} and \cite{gulcicek1987}.

The methodology of this emission model relies on some implicit assumptions that warrant brief discussion. First, the approach assumes that all emissions are optically thin. Lines with the strongest Einstein A coefficients saturate first in their curve of growth and \SI{1304}{\angstrom} has the strongest transition probability herein. The brightness of the HST/COS resolved \SI{1304}{\angstrom} triplet at Ganymede matches the optically thin ratio of 5:3:1 \citep{roth2021_ganymede}. Locally, Ganymede's emissions are bright relative to the other satellites, which assures opacity effects are negligible overall. A second assumption is that electrons remain warm enough to excite all lines at all altitudes; neutral collisions could cool magnetospheric electrons below the various thresholds to excite the different emissions. Cooling rates in Earth's thermosphere are dominated by vibrational excitation of \ce{O2} near \SI{1e8}{cm^{-3}} \citep{Pavlov1999}, a density expected only very near the surface. A third assumption is negligible collisional quenching of long-lived forbidden transitions. With a lifetime of \SI{134}{s} \citep{wiese1996}, \ce{O(^1D)} can be collisionally depopulated before radiating. Thermal \ce{O2} gas quenches \ce{O(^1D)} at a rate of \SI{5e-11}{cm^{3}s^{-1}molecule^{-1}} \citep{Streit1976} and so the critical density where \ce{O2} quenches atoms before radiative decay is \SI{1.5e8}{cm^{3}}--again, a negligible effect over the atmospheric columns.

Table \ref{tbl:rates_combined} gives the modeled emission rate coefficients for the UV and optical emission lines given the relevant electron energies for Europa, Ganymede, and Callisto. Table \ref{tbl:ratios_combined} gives the corresponding emission line ratios for comparison with the observations.

\subsection{Retrievals of atmospheric composition}\label{sec:retrievals}

We use the emission model described in Section \ref{sec:model} to find the best-fit atmospheric column densities for each satellite and date of observation. The model atmosphere is composed of \ce{O}, \ce{O2}, and \ce{H2O}, and the free parameters in the fits are the disk-integrated column densities of each of these constituents. \ce{CO2} is considered but is ultimately not included in the fits because the cross-sections are unavailable at all but one of the optical lines. The modeled auroral emissions are fit to the measurements and upper limits of the four oxygen transitions (at \SIlist[parse-numbers=false]{5577;6300/6364;7774;8446}{\angstrom}) and H$\mathrm{\alpha}$ (\SI{6563}{\angstrom}). An MCMC algorithm is employed to obtain the joint probability distribution of the three atmospheric constituents and hence obtain the most accurate uncertainties on the free parameters. We use the \texttt{emcee} Python implementation \citep{emcee} of the affine-invariant ensemble sampler for Markov chain Monte Carlo (MCMC) described by \cite{goodman2010}. An example output of the MCMC algorithm is shown in Figure \ref{fig:mcmc_example}. In our results we present our final model with uncertainties; upper limits are presented when a species is found to be present at the $<2 \mathrm{\sigma}$ level. In the Appendix we also report the single maximum likelihood model for each observation.

The strongest constraints on atmospheric composition come from simultaneous measurements that cover the largest range of emission lines. The atmospheric retrievals on data from a single observation have the advantage that all measurements are simultaneous. However, numerous measurements have also been made of the UV auroral lines \citep[e.g.,][]{mcgrath2013,roth2016}. We therefore also retrieve the atmospheric composition using the average measurements of auroral emissions at each transition, including past UV measurements.

For the optical transitions, the values in the average fits are averaged over the two dates for each satellite that had good observing conditions and low scattered light (i.e. 1998-11-15 and 2021-06-08 for Ganymede and 2021-05-20 and 2021-06-21 for Europa; see Table \ref{tbl:obs}). For the UV transitions at Europa, we use values of \SIlist{40 \pm 4; 80 \pm 8}{R} for the \SIlist{1304;1356}{\angstrom} emissions respectively, based on 19 observations summarized in \cite{roth2016}. \cite{roth2021_europa} also present measurements of the UV emissions from the subjovian hemisphere in and out of eclipse that are about half the average value cited above. While these data are a closer match to the viewing geometry of our observations, \cite{roth2021_europa} show that the aurora are very similar in and out of eclipse, and we adopt the average values rather than the subjovian values because in the case of Europa the time variability seems to be dominated by the plasma sheet distance and stochastic variability rather than hemispheric trends. Using the UV eclipse measurements instead of the average UV measurements in our modeling provides a much worse fit.\par

The situation is different for Ganymede, where the interactions between Ganymede's magnetic field and Jupiter's magnetosphere result in distinct auroral differences between the hemispheres \citep{mcgrath2013}. In this case, we use the average of the two available subjovian UV measurements, giving values of \SIlist{15 \pm 2; 36 \pm 2}{R} for the \SIlist{1304;1356}{\angstrom} emissions respectively \citep{roth2021_ganymede}. While one of these measurements was made in eclipse and one in sunlight, the values are almost identical between the two. Averaging the leading, trailing, and sub-jovian hemisphere measurements from that work gives higher values of \SIlist{23 \pm 1; 46 \pm 1}{R}, which provide a worse fit when the retrievals are run jointly with the optical data.\par
Ly-$\mathrm{\alpha}$ emission has also been detected on both satellites; on Europa, a value of \SI{70 \pm 26}{R} was found based on two off-limb averages \citep{roth2014_apocenter}. However, the authors note that these emissions may originate from an extended H cloud around Europa rather than directly from \ce{H2O} dissociation. At Ganymede, the \ce{H2O}-sourced Ly-$\mathrm{\alpha}$ emission level predicted by \cite{roth2021_ganymede} of \SI{200}{R} would be inseparable from spatial variations in the surface reflections \citep{roth2021_ganymede,alday2017}. We therefore do not include Ly-$\mathrm{\alpha}$ values in our average fits; including upper limits would not affect the fits because our H$\mathrm{\alpha}$ upper limits provide a much more stringent constraint on \ce{H2O} as a parent molecule because of the smaller uncertainties due to observing in eclipse.
\section{Results \& Discussion}\label{sec:results}\label{sec:disc}

\begin{figure*}
    \centering
    \includegraphics[width=0.7\textwidth]{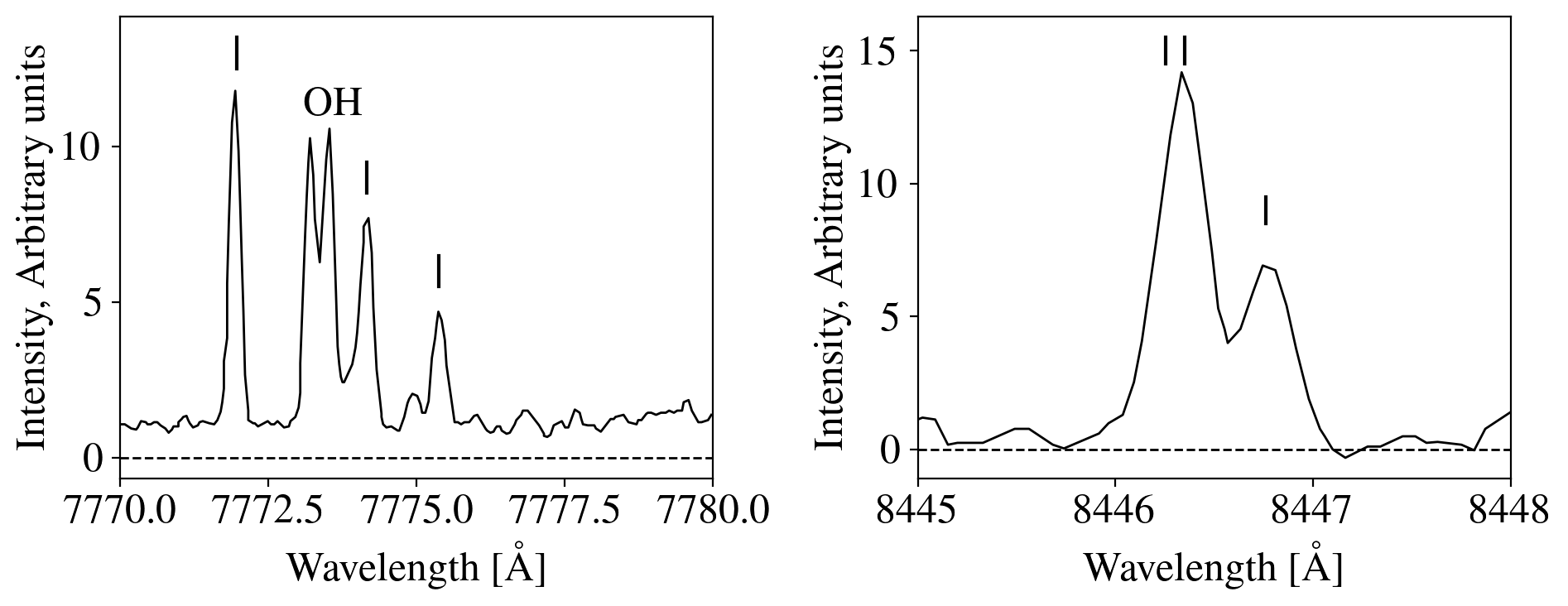}
    \caption{Spectrum covering the \SIlist{7774;8446}{\angstrom} emissions from Earth's atmosphere, obtained with Keck/HIRES. Figure adapted from \cite{slanger2004}.}
    \label{fig:slanger}
\end{figure*}

We report detections of \ce{O}\,I emission at \SIlist[parse-numbers=false]{6300/6364;5577;7774}{\angstrom} at Europa and Ganymede, and additionally \SI{8446}{\angstrom} at Ganymede only, as well as a detection of \SIlist[parse-numbers=false]{6300/6364}{\angstrom} emission at Callisto. These constitute the first detections of these lines at any of the icy Galilean satellites, with the exception of the recent measurements of \SI[parse-numbers=false]{6300/6364}{\angstrom} at Europa \citep{dKB2018,dKB2019}, and also the first detections of \SIlist{7774;8446}{\angstrom} emission at a planetary body other than Earth. An Earth spectrum covering these emissions obtained with Keck/HIRES from \cite{slanger2004} is shown in Figure \ref{fig:slanger} for context. Emission from these lines has also been modeled at Mars and Venus \citep[e.g.][]{bougher2017,borucki1996,gronoff2008}, but is not strong enough to have been detected to date. \SI{7774}{\angstrom} emission was reported in an experiment to detect lightning on Venus \citep{Hansell1995}, but non-detection by a dedicated instrument aboard \textit{Akatsuki} suggests this was spurious \citep{lorenz2019}.

\begin{figure*}
    \centering
    \includegraphics[width=\textwidth]{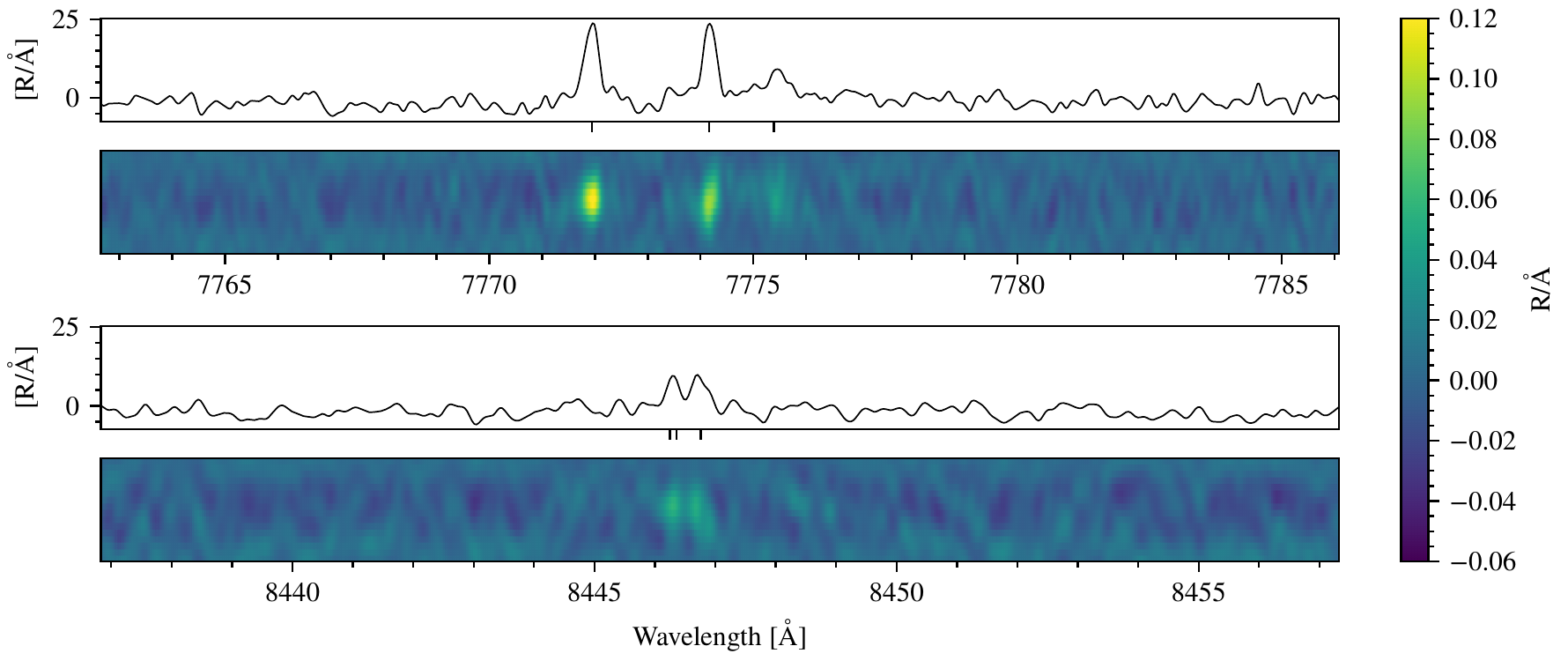}
    \caption{Detections of \SIlist{7774;8446}{\angstrom} \ce{O}\,I emission at Ganymede. For each emission, the lower panel shows a calibrated, background-subtracted image of the spectrum so it can be seen that the emission is localized along the slit (vertical axis), while the upper panel shows the spectrum summed across spatial bins containing the localized emission. The vertical ticks between the panels indicate the locations of the Doppler-shifted emission wavelengths for each transition. Each spectrum has been smoothed with a Gaussian kernel to enhance the features relative to the background.}
    \label{fig:newdets_ganymede}
\end{figure*}

\begin{figure*}
    \centering
    \includegraphics[width=\textwidth]{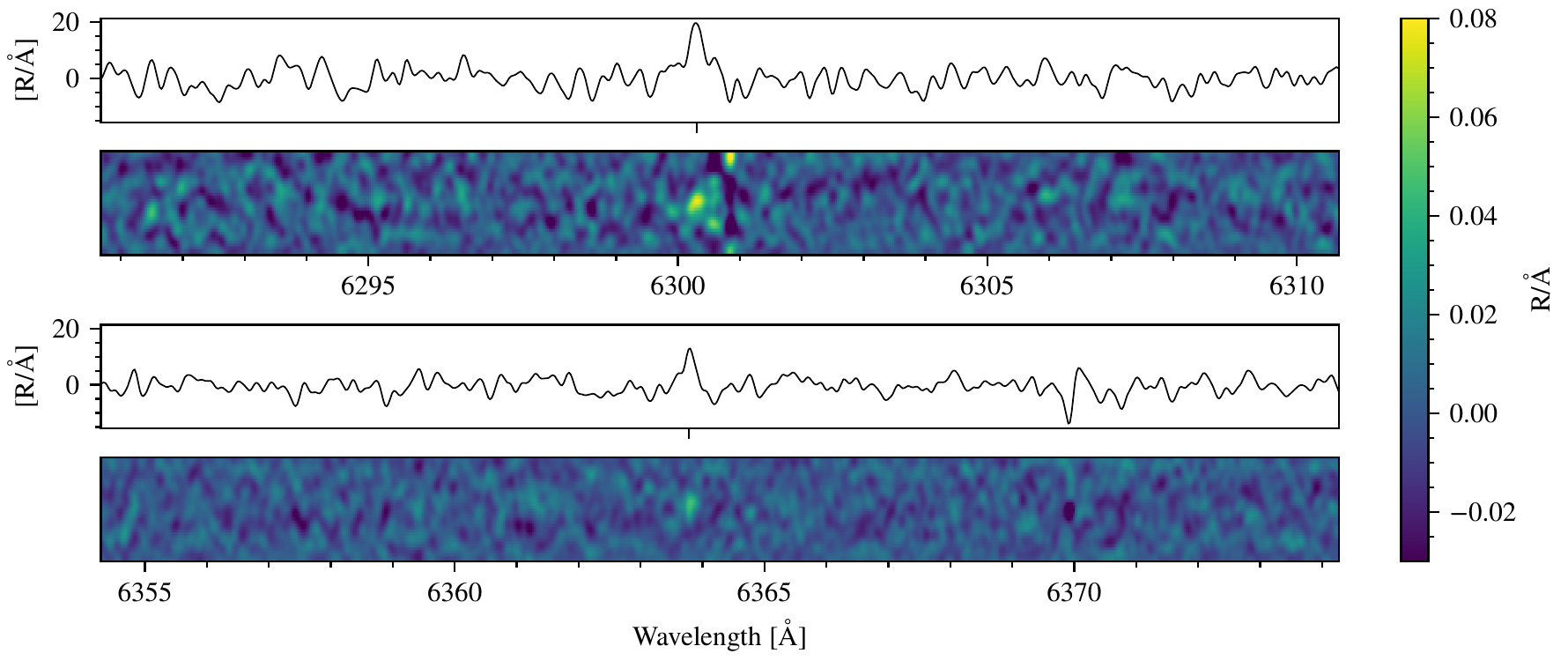}
    \caption{Detections of \SIlist{6300;6364}{\angstrom} \ce{O}\,I emission at Callisto. For each emission, the lower panel shows a calibrated, background-subtracted image of the spectrum so it can be seen that the emission is localized along the slit (vertical axis), while the upper panel shows the spectrum summed across spatial bins containing the localized emission. The vertical ticks between the panels indicate the locations of the Doppler-shifted emission wavelengths for each transition. Each spectrum has been smoothed with a Gaussian kernel to enhance the features relative to the background. In the \SI{6300}{\angstrom} case, the bright points near the top and bottom of the slit just to the right of Callisto are residual from the subtraction of Earth's \SI{6300}{\angstrom}, which is Doppler-shifted relative to the target.
    \label{fig:newdets_callisto}}
\end{figure*}

The \SI[parse-numbers=false]{6300/6364}{\angstrom} emission is seen in every observation of both targets, while \SI{5577}{\angstrom} emission is seen on several but not all dates. \cite{Cassidy2008} proposed electron-excited \ce{Na} could explain Europa's eclipsed emission in the \SIrange{2000}{10500}{\angstrom} wavelength range (clear filter) during \textit{Cassini}'s Jupiter flyby. While the  foreground \ce{Na} nebula from Io hinders our ability measure auroral \ce{Na}, eclipsed \ce{O} emissions are more than an order of magnitude brighter than the \ce{Na} D lines at \SIlist{5890;5896}{\angstrom} and  and thus a more viable explanation for the broadband \textit{Cassini} measurement. The \SIlist{7774;8446}{\angstrom} detections at Ganymede are shown in Figure \ref{fig:newdets_ganymede}. The detection of \SIlist{6300;6364}{\angstrom} emission at Callisto is shown in Figure \ref{fig:newdets_callisto}. The sole previous measurement of Callisto's oxygen atmosphere was attributed to photo-dissociation of \ce{O2} \citep{cunningham2015} and so the emission we report here in shadow constitutes the first definitive detection of Callisto's electron-excited aurora at any wavelength.

Table \ref{tbl:results} provides the disk-averaged surface brightnesses (or 2$\mathrm{\sigma}$ upper limits) of Europa, Ganymede, and Callisto on each observation and for every measured emission line. For poor weather nights, when flux calibration was not reliable, we only report whether or not each line is detected. Table \ref{tbl:BothModels} presents the best-fit atmospheric composition for Europa and Ganymede for each observation, as well as the best-fit composition based on the optical and UV lines averaged across several dates of observation. Note that the exact values of column density are subject to uncertainty due to the poorly-constrained density of the electrons exciting the emissions, which vary by a factor of a few depending on the study \citep[e.g.][]{bagenal2015,bagenal2020,hall1995}. However, the relative contribution from different molecules to the emissions is robust to this uncertainty since it is constrained by the ratio between emission at different wavelengths (given in Table \ref{tbl:ratios_combined}). The modeled and measured brightnesses of all emission lines are shown in Figures \ref{fig:EuropaModel} and \ref{fig:GanymedeModel} and presented in detail in the Appendix in Table \ref{tbl:CombinedModel}.

For Europa, we find that the atmosphere is dominated by \ce{O2} with a column density of \SIrange{3.7}{4.1}{\times 10^{14} cm^{-2}}. The accurate reproduction of several line ratios via dissociative
excitation of \ce{O2} effectively rules out the possibility of an atomic atmosphere proposed by \cite{shemansky2014}, and the column density of atomic oxygen is constrained to be $< \SI{1e13}{cm^{-2}}$. Our models tentatively (2.4$\mathrm{\sigma}$) indicate the presence of \ce{H_2O} at a column density of \SI{1.2\pm0.5e14}{cm^{-2}}. Our derived \ce{O2} column density is within the range found by past UV observations; the average brightness at \SI{1356}{\angstrom} found by \cite{roth2016} corresponds to an average column density of \SI{3.75e14}{cm^{-2}} after converting from the \SI{40}{cm^{-3}} electron density used in their calculation to the \SI{160}{cm^{-3}} used in ours. Our upper limit on \ce{O} of $< \SI{1e13}{cm^{-2}}$ is similarly consistent with the upper limit of \SI{6e12}{cm^{-2}} derived from UV eclipse observations \citep{roth2021_europa}. However, our tentative detection of \ce{H2O} at \SI{1.2 \pm 0.5e14}{cm^{-2}} is a factor of 10 below the estimated \ce{H2O} column density derived from the two UV lines by \cite{roth2021_europa}; this difference is explored in Section \ref{sec:disc_europa_h2o}.

For Ganymede, we find that the atmosphere is dominated by \ce{O2} with a column density of \SIrange{3.2e14}{4.8e14}{cm^{-2}}, and place an upper limit on \ce{O} of \SI{3e13}{cm^{-2}}. This upper limit is consistent with the upper limit of \SI{2e12}{cm^{-2}} derived from UV eclipse observations \citep{roth2021_ganymede}, although the column density we find for \ce{O2} is somewhat higher than the \SI{2.8e14}{cm^{-2}} adopted by these authors in their modeling (using the same electron density as in our model: \SI{20}{cm^{-3}}). We also place an upper limit on \ce{H2O} of \SI{3e13}{cm^{-2}}. This is lower by a factor of $\sim$100 than the \ce{H2O} abundance inferred by \cite{roth2021_ganymede}. Water in Ganymede's atmosphere is discussed in more detail in Section \ref{sec:disc_ganymede_h2o}.

These derived \ce{O2} column densities roughly match predictions by exosphere modeling work for Europa and Ganymede in eclipse \citep{oza2019,leblanc2017}, although we note that the model predictions are for zenith columns whereas our observations are disk-integrated and likely dominated by the tangent column at the limb.

The best-fit models for individual nights occasionally find a contribution from \ce{O} or \ce{H2O}, which could be physically meaningful or could simply be a reflection of the high uncertainty arising from performing retrievals on low signal-to-noise data from individual eclipses. For Ganymede on 2018-06-15, the sky background level was anomalously high (see Table \ref{tbl:obs}) and the inference of \ce{H2O} should be viewed with skepticism despite the high mathematical significance. For Europa on 2021-06-21, the contribution from O is marginal (at the 2.3$\mathrm{\sigma}$ level), though the data quality is good from this night.

For Callisto, the detection of only one excited \ce{O}\,I state (\SI[parse-numbers=false]{6300/6364}{\angstrom}), combined with the lack of past electron-excited UV detections, limits the complexity of models that can be fit to the data. Using a model that assumes a pure \ce{O2} atmosphere and an electron density at Callisto of \SI{0.15}{cm^{-3}}, we find a column density of \SI{4.0 \pm 0.9e15}{cm^{-2}}, which matches very well with the \SI{4.0e15}{cm^{-2}} inferred from a UV measurements for which the excitation was attributed to photoelectrons \citep{cunningham2015}. Callisto will be discussed further in Section \ref{sec:disc_callisto}.

\begin{figure*}
    \centering
    \includegraphics[width=0.9\textwidth]{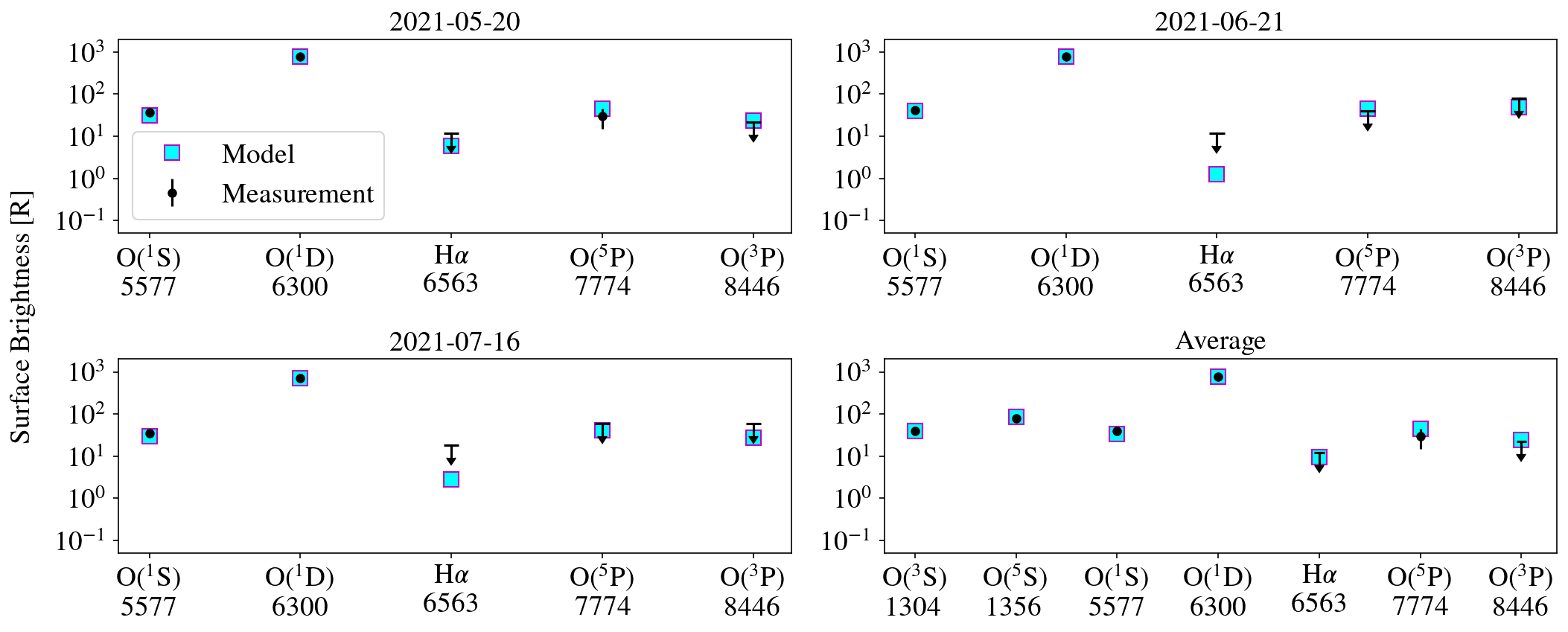}
    \caption{Measured auroral emissions along with the best-fit atmospheric model for Europa for each date of observation and for the average auroral emissions including past UV measurements \citep{roth2016}.}
    \label{fig:EuropaModel}
\end{figure*}

\begin{figure*}
    \centering
    \includegraphics[width=0.9\textwidth]{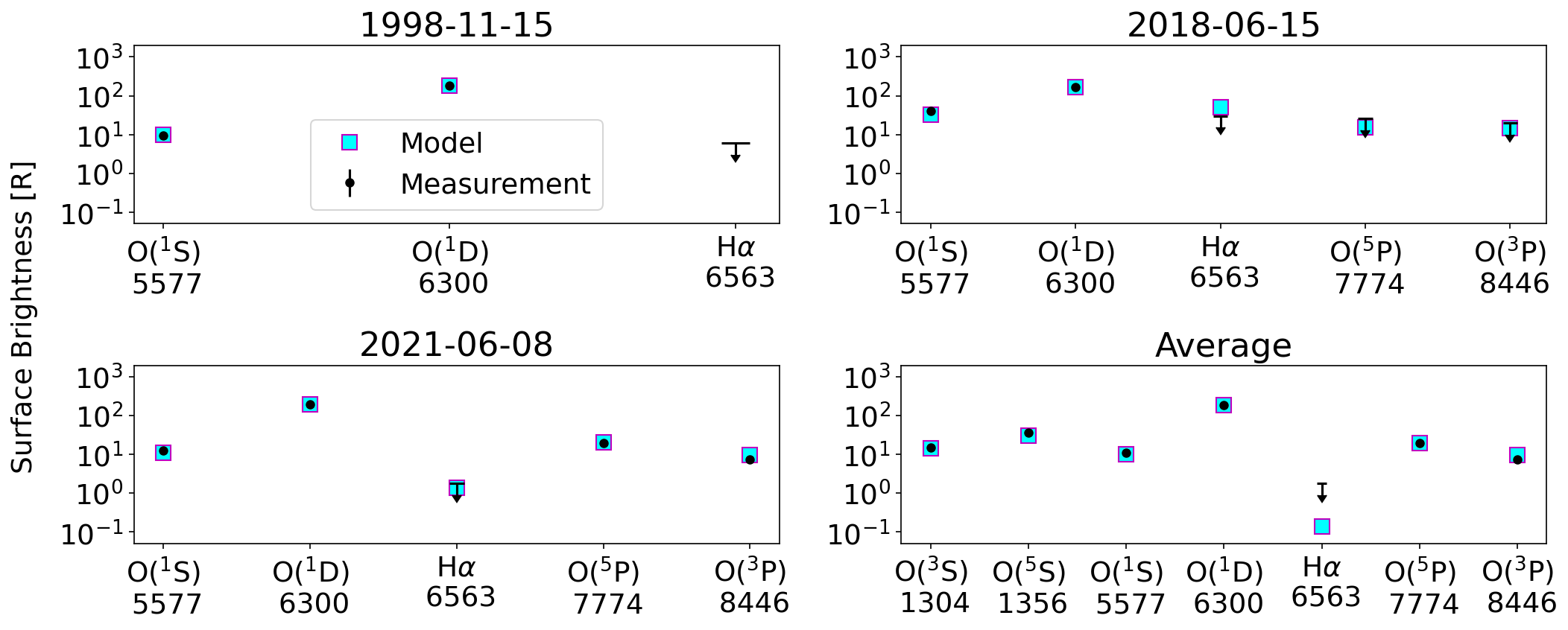}
    \caption{Measured auroral emissions along with the best-fit atmospheric model for Ganymede for each date of observation and for the average auroral emissions including past UV measurements \citep{roth2021_ganymede}.}
    \label{fig:GanymedeModel}
\end{figure*}

\begin{deluxetable*}{lccccccc}
\tablecaption{Measured auroral surface brightnesses and upper limits [R].\textsuperscript{b} \label{tbl:results}}
\tablecolumns{8}
\tablewidth{0pt}
\centering
\tablehead{
 & \colhead{\SI{6300}{\angstrom}} & \colhead{\SI{6364}{\angstrom}} & \colhead{\SI{5577}{\angstrom}} & \colhead{\SI{7774}{\angstrom}} & \colhead{\SI{8446}{\angstrom}} & \colhead{\SI{6563}{\angstrom}}\\[-5pt]
\colhead{Satellite/Date} & \colhead{\ce{O(^1D)}} & \colhead{\ce{O(^1D)}} & \colhead{\ce{O(^1S)}} & \colhead{\ce{O({3p}^5P)}} & \colhead{\ce{O({3p}^3P)}} & \colhead{H$\mathrm{\alpha}$}
}
\startdata
\textbf{Ganymede} & \\ 
\quad 1998-11-15 & \num{139 \pm 3} & \num{42.9 \pm 1.7} & \num{9.2 \pm 1.6} & --- & --- & $< 6$\\
\quad 2018-06-15 & \num{127 \pm 7} & \num{36 \pm 5} & \num{41 \pm 4} & $< 26$ & $< 20$ & $< 30$\\
\quad 2021-06-08 & \num{147.6 \pm 1.5} & \num{45.0 \pm 1.0} & \num{12.5 \pm 0.6} & \num{20 \pm 2} & \num{7.3 \pm 1.4} & $< 1.8$\\
\quad 2021-10-01\textsuperscript{c} & \textit{detected} & \textit{detected} & --- & --- & --- & ---\\
\quad Average & $\mathit{143 \pm 2}$ & $\mathit{44 \pm 1}$ & $\mathit{11 \pm 1}$ & $\mathit{20 \pm 2}$ & $\mathit{7.3 \pm 1.4}$ & ---\\
\hline
\textbf{Europa} & \\
\quad 2018-03-22\textsuperscript{a,c} & \textit{detected} & \textit{detected} & --- & --- & --- & --- \\
\quad 2021-05-20 & \num{573 \pm 14} & \num{190 \pm 7} & \num{37 \pm 4} & \num{30 \pm 15} & $< 22$ & $< 12$\\
\quad 2021-06-21 & \num{613 \pm 12} & \num{176 \pm 6} & \num{42 \pm 5} & $< 40$ & $< 80$ & $< 12$\\
\quad 2021-07-16 & \num{547 \pm 14} & \num{170 \pm 8} & \num{35 \pm 8} & $< 60$ & $< 60$ & $< 18$\\
\quad Average & $\mathit{593 \pm 9}$ & $\mathit{183 \pm 4}$ & $\mathit{40 \pm 3}$ & $\mathit{30 \pm 15}$ & --- & ---\\
\hline
\textbf{Callisto} & \\
\quad 2021-07-04 & \num{6.1 \pm 1.7} & \num{3.3 \pm 1.2} & $< 2$ & $< 8$ & $< 4$ & $< 3$\\
\quad 2021-09-26\textsuperscript{c} & --- & --- & --- & --- & --- & ---\\
\enddata
\vspace*{3pt}
\textsuperscript{a}Previously published in \cite{dKB2018}.\\
\textsuperscript{b}2$\mathrm{\sigma}$ upper limits; we require a measurement to be at or above the 2$\mathrm{\sigma}$ level to claim a detection.\\
\textsuperscript{c}On poor weather nights when it was not possible to flux-calibrate the data, we report only whether a given line is detected.
\end{deluxetable*}

\begin{deluxetable}{lccc}
\tablecaption{Best-fit model atmospheres\textsuperscript{a}.\label{tbl:BothModels}}
\tablecolumns{4}
\tablewidth{0pt}
\centering
\tablehead{
 & \multicolumn{3}{c}{Column Density [$\times\, 10^{14}\,\si{cm^{-2}}$]}\\[3pt]
\cline{2-4}
\colhead{Satellite/Date} & \colhead{\ce{O2}} & \colhead{\ce{O}} & \colhead{\ce{H2O}}
}
\startdata
\textbf{Europa}\\
\quad 2021-05-20 & \num{4.0 \pm 0.1} & $< 0.4$ & $< 0.9$\\
\quad 2021-06-21 & \num{3.7 \pm 0.2} & \num{1.8 \pm 0.8} & $< 0.8$\\
\quad 2021-07-16 & \num{3.7 \pm 0.2} & $< 1.2$ & $< 1.1$ \\
\quad Average\textsuperscript{b} & \num{4.1 \pm 0.1} & $< 0.1$ & \num{1.2 \pm 0.5}\\
\hline
\textbf{Ganymede}\\
\quad 1998-11-15 & \num{4.5 \pm 0.1} & $< 3$ & $< 0.6$\\
\quad 2018-06-15 & \num{3.2 \pm 0.3} & $< 8$ & \num{16 \pm 3}\\
\quad 2021-06-08 & \num{4.8 \pm 0.1} & $< 0.8$ & $< 0.5$\\
\quad Average\textsuperscript{b} & \num{4.7 \pm 0.1} & $< 0.3$ & $< 0.3$\\
\hline
\textbf{Callisto}\\
\quad 2021-07-04 & \num{40 \pm 9} & --- & --- \\
\enddata
\vspace*{3pt}
\footnotesize{\textsuperscript{a}For the following assumed electron parameters, as described and referenced in the text: Europa $n_e=\SI{160}{cm^{-3}}$ with 95\% Maxwellian-distributed in energy about \SI{20}{eV} and 5\% about \SI{250}{eV}; Ganymede $n_e=\SI{20}{cm^{-3}}$ Maxwellian-distributed about \SI{100}{eV}; Callisto $n_e=\SI{0.15}{cm^{-3}}$ Maxwellian-distributed about \SI{35}{eV}.}\\
\footnotesize{\textsuperscript{b}Average atmospheric compositions are based on fits to the average optical and UV brightnesses across all available dates of observation, as described in Section \ref{sec:retrievals}.}
\end{deluxetable}

\subsection{\texorpdfstring{\ce{H2O}}{H2O}-dominated atmospheres on Europa and Ganymede?}\label{sec:disc_europa_h2o}\label{sec:disc_ganymede_h2o}

Recently, \cite{roth2021_europa} and \cite{roth2021_ganymede} postulated \ce{H2O}-dominated atmospheres on the trailing hemispheres on Europa and Ganymede. This claim is based in part on an observed \SI[parse-numbers=false]{1356/1304}{\angstrom} ratio that is lower for both satellites on the trailing hemisphere than on the leading hemisphere. While this difference had previously been attributed to a greater \ce{O} abundance on the trailing hemisphere \citep{roth2016,molyneux2018}, the new observations included eclipse measurements of the subjovian hemisphere, permitting a stronger upper limit on the column density of \ce{O} by placing a direct constraint on the contribution to \SI{1304}{\angstrom} from resonant scattering by \ce{O}. 

Our observations are all made in eclipse and therefore image the subjovian hemisphere. In the case of Europa, our upper limit for \ce{O} of \SI{\sim1e13}{cm^{-2}} (see Table \ref{tbl:results}) is consistent with the upper limit of \SI{6e12}{cm^{-2}} derived by \cite{roth2021_europa}. However, our best-fit value for \ce{H2O}, \SI{1.2\pm0.5e14}{cm^{-2}}, is a factor of 10 lower than the disk-averaged \ce{H2O} abundance of \SI{\sim1e15}{cm^{-2}} found by \cite{roth2021_europa} for the sunlit trailing hemisphere. Note that while the column densities presented here depend on the assumed electron properties, we have adopted the same electron properties as used by \cite{roth2021_europa} and \cite{roth2021_ganymede}, so the factors by which the derived column densities differ between studies are not sensitive to uncertainty in electron density and are only weakly sensitive to uncertainty in electron temperature. Similarly, while the absolute abundances of species derived from our modeling depends on electron density, the relative abundances do not and are therefore not subject to uncertainty on this parameter.

The presence of \ce{H2O} in our modeling is only indicated at the 2.4$\mathrm{\sigma}$ level and only after averaging datasets together with the UV; for individual nights, we find only upper limits on \ce{H2O}, in the \SIrange{0.8e14}{1.1e14}{cm^{-2}} range. For all UV and optical oxygen lines except \SI{5577}{\angstrom}, the emission rates are at least $10\times$ higher for \ce{O2} than for \ce{H2O}, so that these lines are not strongly sensitive to \ce{H2O} (see Table \ref{tbl:rates_combined}). However, for \SI{5577}{\angstrom} the emission rates are comparable between the species. This line, combined with H$\mathrm{\alpha}$, therefore provide the strongest constraints on water abundance. In particular, the \ce{H2O} disk-integrated column density from \cite{roth2021_europa} would produce \SI{70}{R} of emission at H$\mathrm{\alpha}$, compared to the 2$\mathrm{\sigma}$ upper limit of \SI{12}{R} from our Europa observations, in addition to predicting \ce{O}\,I line ratios inconsistent with the optical and UV lines.

To determine whether an atmosphere containing \ce{O}, \ce{O2}, \ce{H2O} is preferred over an \ce{O2}-only atmosphere, we fit the data using both the three-species model and the \ce{O2}-only model and find that the fit is only marginally better when all three species are included.
Given the lack of strong preference for the \ce{H2O}-containing model and the fact that \ce{H2O} is only preferred by our average fits (and only at 2.4$\mathrm{\sigma}$), and not on individual nights, we consider the detection of Europa's \ce{H2O} at all to be tentative in our data. Moreover, we strongly rule out a water abundance of \SI{2.5e14}{cm^{-2}} or higher, or alternatively an \ce{H2O}/\ce{O2} ratio above 0.5, on the eclipsed subjovian hemisphere. 

For Ganymede, our observations on the two nights that had clear sky conditions as well as low Jupiter scattered light conditions (see Table \ref{tbl:obs}) place an upper limits on \ce{H2O} of \SI{3e13}{cm^{-2}}, or a maximum \ce{H2O}/\ce{O2} ratio of 0.06. In contrast, the modeled ratio for the center trailing hemisphere from the UV data was found to be 12--32 and for the leading hemisphere was found to be 2--5 \citep{roth2021_ganymede}, with the disk-averaged ratios a factor of a few lower. This \ce{H2O} abundance contributed \SI{1}{R} to the disk-averaged \SI{1304}{\angstrom} emission but should contribute \SI{23}{R} and \SI{46}{R} to the optical emission at \SI{5577}{\angstrom} and \SI{6563}{\angstrom} respectively, both of which are strongly ruled out in our observations.

There are several differences between the optical and UV observations that may be relevant to these different results, in particular the low \ce{H2O/O2} ratios ($<$1) derived from our optical eclipse data of both satellites, and the much higher ratios of 10--30 determined for the trailing hemispheres in the UV.

First, our observations were made with the targets in shadow while the UV observations were made with the targets in sunlight. For an \ce{H2O} atmosphere sustained by sublimation, the column density should be lower in eclipse due to the lower surface temperature. \cite{leblanc2017} modeled Ganymede's \ce{H2O} exosphere in eclipse and calculated a 4 orders-of-magnitude collapse, from a peak column density in the $10^{14}$  to $10^{16}$ cm$^{-2}$ range. However, in such a case the UV auroral brightnesses would also be lower in eclipse than in sunlight, whereas the observed brightnesses of both satellites' aurora are comparable in and out of eclipse \citep{roth2021_europa,roth2021_ganymede}. A sublimation atmosphere that partially collapses in eclipse therefore cannot reconcile the UV and optical datasets. For Europa, even the \ce{O2} atmosphere is modeled to drop by an order of magnitude in eclipse \citep{oza2019}, which is not consistent with the UV observations either.

The observations also differ in that we target the subjovian hemisphere, while the UV-derived \ce{H2O} abundances were highest on the trailing hemispheres of both Europa and Ganymede. For Europa, the UV line ratio is also just as low, and hence as inconsistent with pure \ce{O2}, on the subjovian hemisphere as it is on the trailing hemisphere \citep{roth2021_europa}. The past UV data and our optical results together are therefore hard to reconcile with the presence of significant \ce{O} or \ce{H2O} on Europa, which seems to require a new explanation for the low \SI[parse-numbers=false]{1356/1304}{\angstrom} ratio at least on the subjovian hemisphere.

For Ganymede, the UV line ratio is higher on the subjovian hemisphere than either leading or trailing \citep{roth2021_ganymede} and is consistent with pure \ce{O2}. The difference between the UV and optical data for Ganymede can thus be attributed to an \ce{H2O} atmosphere that is only present on the leading/trailing hemispheres, and not on the subjovian, regardless of whether it is sunlit or eclipsed.

Given the clear differences between hemispheres on both moons, it may be that the upper limit derived for \ce{O} on the subjovian hemispheres is not applicable to the trailing hemispheres, and hence that both \ce{O} and \ce{H2O} remain candidates to explain the low UV ratio on the trailing hemispheres. If atomic \ce{O} is primarily a  product of electron-impact dissociation of \ce{O2}, it is similarly predicted to be most abundant on the center of the trailing hemisphere for Europa \citep{cassidy2013}, so either species would produce a \SI[parse-numbers=false]{1356/1304}{\angstrom} ratio that increases from center to limb of the trailing hemisphere as observed.

A sputtered \ce{H2O} atmosphere has also been proposed for both satellites \citep{johnson1981}. For Ganymede, recent modeled sputtered \ce{H2O} column densities are in the $10^{11}$  to $10^{13}$ cm$^{-2}$  range, while modeled sublimated column densities are on the order of $\sim10^{16}$ cm$^{-2}$  \citep{marconi2007,leblanc2017,vorburger2022}. Such sublimated column densities are ruled out in our data, but the expected sputtered column densities are low enough to be below our detection limits. For Europa, sublimated \ce{H2O} should be roughly two orders of magnitude lower than for Ganymede based on their surface temperatures, while sputtered \ce{H2O} is limited to $10^{13}$ cm$^{-2}$  as for Ganymede \citep{feistel2007,plainaki2013,shematovich2005,smyth2006}. Both predictions are near our detection limits, so we cannot provide strong constraints on a potential sputtered \ce{H2O} atmosphere. The expected sputtering production is also longitudinally variable. For Ganymede, \cite{leblanc2017} show that the \ce{H2O} sputtering production should be highest on the leading hemisphere due to the magnetic field geometry, while for Europa \cite{cassidy2013} show that sputtering rate of all species is highest on the trailing hemisphere.

\subsection{Time-variability of Europa's aurora}\label{sec:disc_europa_scaleheight}

While Ganymede's aurora are produced by complex interactions between the incoming plasma and Ganymede's magnetic field, Europa's auroral brightness should be a more straight-forward product of the local electron density and the column density of species in its atmosphere. In the UV, the aurora are found to vary by a factor of 5--10 across 71 total observations \citep{roth2016}; there is a correlation with Europa's distance from the jovian plasma sheet, although there is significant variability between individual exposures as well as between observations at similar plasma sheet locations. There is no indication that there is a difference in brightness between sunlight and eclipse.

\begin{figure}
    \centering
    \includegraphics[width=\columnwidth]{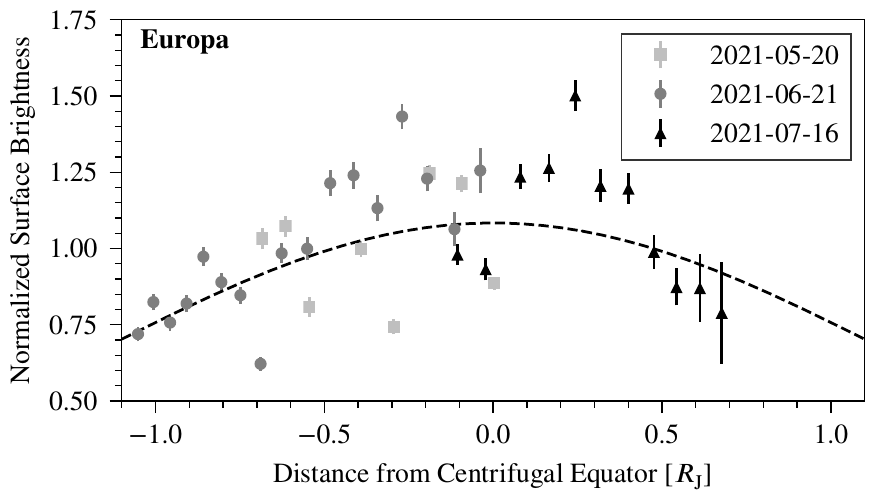}
    \caption{Brightness of Europa's aurora as a function of plasma sheet distance during each observation. Each observation has been normalized to its average. The dashed curve corresponds to a scale height of 1.67\,$R_\text{J}$.}
    \label{fig:Europa_plasmasheet}
\end{figure}

Our individual integrations have sufficient signal-to-noise in the \SI[parse-numbers=false]{6300/6364}{\angstrom} emission that we can analyze the changes in the aurora through each eclipse, during which the plasma sheet distance is changing but longer-timescale variations in either Europa's atmosphere or the plasma conditions do not factor in. Figure \ref{fig:Europa_plasmasheet} shows the brightness of the \ce{O(^1D)} transition, the sum of the \SIlist{6300;6364}{\angstrom} lines, as a function of plasma sheet distance \citep[as defined in][]{phipps2021} during each observation. There is an apparent correlation with plasma sheet distance on 2021-07-16 and 2021-06-21, although we note that there is no apparent correlation on 2021-05-20, and even when present the scatter is significant and the correlation reverses slightly right at the equator. We fit a model of the form $B(z)=B_0e^{-(z/H)^2}$ where $B_0$ is the brightness as Europa is crossing the centrifugal equator of the plasma sheet \citep{phipps2021}; $z$ is the absolute-value distance from plasma sheet in $R_J$, and $H$ is the scale height \citep{hill1974,roth2016}. The free parameters are $B_0$ and $H$, and the best-fit model is shown on Figure \ref{fig:Europa_plasmasheet}. The best-fit scale height of 1.67\,$R_\text{J}$ is in the vicinity of the scale height of 1.6\,$R_\text{J}$ found for the UV aurora \citep{roth2016}, and the plasma torus scale height at Europa's orbit of 1.7\,$R_\text{J}$ from the model of \cite{bagenal2011}. 

The UV and optical observations thus collectively demonstrate that Europa's auroral brightness correlates with magnetic latitude, although there is large scatter about this correlation and other sources of variability may dominate on timescales of minutes to days. We note that Io's aurora also show a factor of $\sim$3 variability even at a given plasma sheet location \citep{oliversen2001,roth2014,schmidt2022}, and that this is indeed expected based on observed variations in the electron density \citep{bagenal2015}.

\subsection{Callisto}\label{sec:disc_callisto}
Callisto's atmosphere is the most poorly understood of the Galilean satellites. The presence of an ionosphere at Callisto was indicated by \textit{Galileo} from the plasma density measured during Callisto flyby \citep{gurnett2000} and from radio occultation observations \citep{kliore2002}. Under the assumption of a predominantly O$_2$ atmosphere (which is supported by modeling work by \cite{liang2005}), the radio occultation measurements were use to infer an \ce{O2} column density of \SI{4e16}{cm^{-2}} \citep{kliore2002}. However, the ionosphere was only detected on the sunlit trailing hemisphere, which is the side impacted by Jupiter's co-rotating plasma, and not on the sunlit leading hemisphere. It was also denser on the sunlit than on the night side of Callisto's terminator, suggesting that photons play a role in producing it. \cite{kliore2002} hypothesized that the atmosphere is generated by sputtering, but it's been suggested that the presence of the ionosphere should divert plasma around Callisto \citep{strobel2002}. If this is the case, it may be that sputtering on the night side generates the atmosphere, which becomes ionized by photo-electrons when it moves into sunlight.

The first detection of Callisto's aurora were made with HST/COS of the \SIlist{1304;1356}{\angstrom} emission lines, in sunlight on the leading/Jupiter-facing hemisphere \citep{cunningham2015}. The observed brightnesses of \SIlist{3.3 \pm 2.8; 3.2 \pm 1.6}{R} for the two emissions respectively correspond to an \ce{O2} column density of \SI{4e15}{cm^{-2}} assuming excitation by photo-electrons. The authors argue that photo- rather than magnetospheric electrons are exciting the emissions, on the basis of the fact that photo-electron-excited emission is modeled to be $10\times$ brighter, assuming the magnetospheric electrons penetrate to the atmosphere at all instead of being diverted around Callisto \citep{cunningham2015}. This atmospheric density is $5\times$ higher than the \ce{CO2} column density of \SI{8e14}{cm^{-2}} derived from a \textit{Galileo} off-limb airglow measurement \citep{carlson1999}, but $10\times$ lower than the trailing hemisphere \ce{O2} density inferred from the ionospheric measurements. 

We present the first detection of Callisto's optical aurora from 2021-07-04, specifically the \ce{O}\,I emissions at \SIlist{6300;6364}{\angstrom} with brightnesses of \SI{6.1 \pm 1.7}{R} and \SI{3.3 \pm 1.2}{R} respectively. Assuming a purely \ce{O2} atmosphere, these emissions correspond to a column density of \SI[parse-numbers=false]{(4.0 \pm 0.9)\times 10^{15}}{cm^{-2}}, which is identical to the column density inferred from past UV observations, although our derived column density depends on the assumed electron energies and densities at Callisto, which are poorly constrained at the current time and likely variable. The UV observations were made in sunlight, and the derived column density were based on the assumption that the aurora were primarily excited by photo-electrons rather than magnetospheric electrons. Our observations were made in eclipse and the detected emissions therefore must be excited by magnetospheric electrons, removing this ambiguity in interpretation. The match between our column density and that inferred from the past UV data may support the interpretation of the excitation mechanism for the UV aurora. However, the large uncertainty on the electron density, and the unknown variability of Callisto's atmosphere and the electron density at Callisto's orbit, prevent firm conclusions. Assuming a purely \ce{O2} atmosphere and excitation by magnetospheric electrons alone, our measured optical emissions would correspond to emission of \SI{0.5}{R} at \SI{1304}{\angstrom} and \SI{1.2}{R} at \SI{1356}{\angstrom} (see Table \ref{tbl:CombinedModel}). These values are lower than the UV aurora measurements of \SI{3.3 \pm 2.8}{R} and \SI{3.2 \pm 1.6}{R}, which suggests photoelectrons are indeed needed to excite the observed UV emissions, although the measurements are still consistent within 1--2$\mathrm{\sigma}$ with magnetospheric electron excitation due to the large uncertainties. In addition, the column densities inferred from aurora for the leading/subjovian hemispheres are an order of magnitude lower than the estimate from radio occultation for the trailing hemisphere of \SI{4e16}{cm^{-2}} \citep{cunningham2015,kliore2002}, which could be due to enhanced atmospheric generation via sputtering on the trailing hemisphere if confirmed by measurements of both hemispheres using the same technique.

The detection of combined \SIlist{6300;6364}{\angstrom} emission at \SI{9.4}{R} along with the upper limit on \SI{5577}{\angstrom} of \SI{2}{R} places a lower limit on the \ce{O(^1D)}/\ce{O(^1S)} ratio of 4.75. This rules out \ce{H2O} as the dominant parent molecule, as it would produce an emission ratio of 1.2 (see Table \ref{tbl:ratios_combined}). Atomic \ce{O} was previously ruled out as the dominant parent species due to an early non-detection of \SI{1304}{\angstrom}, which is produced in part by resonant scattering \citep{strobel2002}. \ce{CO2} is known to be present in Callisto's atmosphere, but the \ce{CO2} column density of \SI{1.1e16}{cm^{-2}} that would be required to produce our measured auroral emissions is two orders of magnitude higher than the detected \ce{CO2} atmosphere \citep{carlson1999}. We conclude that molecular \ce{O2} is the most plausible parent molecule for the aurora we observe at Callisto.

\section{Conclusions}\label{sec:conc}

We present new detections of oxygen auroral emissions at the icy Galilean satellites Europa, Ganymede and Callisto. At Europa we detect emission at \SIlist[parse-numbers=false]{6300/6364;5577;7774}{\angstrom}, including the first detection of emissions at the latter two of these wavelengths. At Ganymede we present the first detections of aurora at any optical wavelength, including measurements at \SIlist[parse-numbers=false]{6300/6364;5577;7774;8446}{\angstrom}. At Callisto we present the first detection at any wavelength of aurora that must be excited by magnetospheric electrons, via a detection of the \SI[parse-numbers=false]{6300/6364}{\angstrom} lines. Upper limits are presented for oxygen lines when not detected, and for hydrogen H$\mathrm{\alpha}$ in all observations.

The emissions from Europa and Ganymede are fit with atmospheres that are permitted to contain \ce{O}, \ce{O2} and \ce{H2O}. The best-fit atmosphere for Europa is predominantly \ce{O2} with a column density of \SI{4.1 \pm 0.1e14}{cm^{-2}}, and we find weak evidence for \ce{H2O} at a column density of \SI{1.2 \pm 0.5e14}{cm^{-2}}. The best-fit atmosphere for Ganymede is exclusively \ce{O2} at a column density of \SI{4.7 \pm 0.1e14}{cm^{-2}}; an upper limit of \SI{3e13}{cm^{-2}} is placed on both atomic \ce{O} and \ce{H2O}. These data place strong constraints on \ce{H2O} abundance because of the unique sensitivity of the \SI{5577}{\angstrom} and H$\mathrm{\alpha}$ emissions to the presence of water.

Callisto's aurora indicate an \ce{O2} column density of \SI{4.0 \pm 0.9e15}{cm^{-2}} for our adopted electron properties. This matches the column density inferred from past UV observations \citep{cunningham2015}, even though the UV aurora were attributed to photoelectron excitation whereas we observe emissions in eclipse that must be excited by magnetospheric electrons. 

These data collectively demonstrate the power of the optical aurora in providing a collection of emission lines that can clearly differentiate between parent species and provide robust constraints on atmospheric make-up. The recently-launched \textit{James Webb Space Telescope} will cover the wavelengths of the majority of the transitions presented here and may provide new insight into not just the strength of the aurora but their spatial variations across the satellites.\\ \\

Support for this work was provided by NASA through
grant to program HST-GO-15425 from the Space Telescope Science Institute, which is operated by the Associations
of Universities for Research in Astronomy, Incorporated, under NASA contract NAS5-26555. We also gratefully acknowledge support from the NASA Solar System Observations program via grants 80NSSC22K0954 and 80NSSC21K1138. The data presented herein were obtained at the W. M. Keck Observatory, which is operated as a scientific partnership among the California Institute of Technology, the University of California and the National Aeronautics and Space Administration. The Observatory was made possible by the generous financial support of the W.~M.~Keck Foundation. The authors wish to recognize and acknowledge the very significant cultural role and reverence that the summit of Maunakea has always had within the indigenous Hawaiian community.  We are most fortunate to have the opportunity to conduct observations from this mountain.

\vspace{5mm}
\facilities{Keck/HIRES}

\clearpage
\appendix

\section{Emission model}

\setlength{\tabcolsep}{2pt}
\begin{deluxetable}{@{\extracolsep{4pt}}lc|cccc|cccc|cccc@{}}
\tablecaption{Emission rate coefficients for Europa, Ganymede and Callisto.\label{tbl:rates_combined}}
\centering
\tablehead{
\colhead{} & \colhead{} & \multicolumn{12}{c}{Parent Species [$\times 10^{-10}$\,\si{cm^3.s^{-1}}]}\\
\cline{3-14}
\colhead{} & \colhead{} & \multicolumn{4}{c}{Europa} & \multicolumn{4}{c}{Ganymede} & \multicolumn{4}{c}{Callisto}\\
\cline{3-6}\cline{7-10}\cline{11-14}
 & \colhead{Wavelength [\si{\angstrom}]} & \colhead{\ce{O}} & \colhead{\ce{O2}} & \colhead{\ce{H2O}} & \colhead{\ce{CO2}} & \colhead{\ce{O}} & \colhead{\ce{O2}} & \colhead{\ce{H2O}} & \colhead{\ce{CO2}} & \colhead{\ce{O}} & \colhead{\ce{O2}} & \colhead{\ce{H2O}} & \colhead{\ce{CO2}}
}
\startdata
\ce{O(^1S)} & 5577 & \num{4.8} & \num{4.4} & \num{2.5} & \num{40} & \num{2.4} & \num{11} & \num{7.6} & \num{85} & \num{4.4} & \num{6.7} & \num{4.2} & \num{57} \\
\ce{O(^1D)} & $6300+6364$ & \num{33} & \num{118} & \num{3.1} & --- & \num{7.9} & \num{200} & \num{9.4} & --- & \num{23} & \num{156} & \num{5.2} & --- \\
\ce{O(^3S)} & 1304 & \num{37} & \num{5.9} & \num{0.33} & \num{1.1} & \num{50} & \num{14} & \num{1.3} & \num{3.8} & \num{37} & \num{5.9} & \num{0.3} & \num{1.1} \\
\ce{O(^5S)} & 1356 & \num{4.6} & \num{13} & \num{0.082} & \num{1.0} & \num{1.1} & \num{32} & \num{0.33} & \num{3.9} & \num{4.6} & \num{13} & \num{0.08} & \num{1.0} \\
\ce{O({3p}^3P)} & 8446 & \num{9.8} & \num{3.6} & \num{0.40} & --- & \num{7.2} & \num{9.9} & \num{1.4} & --- & \num{9.8} & \num{3.6} & \num{0.4} & --- \\
\ce{O({3p}^5P)} & 7774 & \num{2.0} & \num{6.8} & \num{0.20} & --- & \num{0.54} & \num{21} & \num{0.53} & --- & \num{2.0} & \num{6.8} & \num{0.2} & --- \\\
Ly-$\mathrm{\alpha}$ & 1216 & --- & --- & \num{9.3} & --- & --- & --- & \num{33} & --- & --- & --- & \num{9.3} & ---  \\
H$\mathrm{\alpha}$ & 6563 & --- & --- & \num{5.0} & --- & --- & --- & \num{15} & --- & --- & --- & \num{5.0} & --- \\
\enddata
\end{deluxetable}

\begin{deluxetable}{@{\extracolsep{4pt}}lc|cccc|cccc|cccc@{}}
\tablecaption{Modeled emission ratios for Europa, Ganymede and Callisto.\label{tbl:ratios_combined}}
\centering
\tablehead{
\colhead{} & \colhead{} & \multicolumn{12}{c}{Parent Species}\\
\cline{3-14}
\colhead{} & \colhead{} & \multicolumn{4}{c}{Europa} & \multicolumn{4}{c}{Ganymede} & \multicolumn{4}{c}{Callisto}\\
\cline{3-6}\cline{7-10}\cline{11-14}
Ratio & \colhead{Wavelengths [\si{\angstrom}]} & \colhead{\ce{O}} & \colhead{\ce{O2}} & \colhead{\ce{H2O}} & \colhead{\ce{CO2}} & \colhead{\ce{O}} & \colhead{\ce{O2}} & \colhead{\ce{H2O}} & \colhead{\ce{CO2}} & \colhead{\ce{O}} & \colhead{\ce{O2}} & \colhead{\ce{H2O}} & \colhead{\ce{CO2}}
}
\startdata
$\text{Ly-}\mathrm{\alpha}/\text{H-}\mathrm{\alpha}$ & $1216/6563$ & --- & --- & 1.9 & --- & --- & --- & 2.1 & --- & --- & --- & 2.0 & --- \\
$\ce{O(^1D)}/\text{H-}\mathrm{\alpha}$ & $(6300+6364)/6563$ & --- & --- & 0.61 & --- & --- & --- & 0.61 & --- & --- & --- & 0.60 & --- \\
$\ce{O(^1D)}/\ce{O(^1S)}$ & $(6300+6364)/5577$ & 6.7 & 27 & 1.2 & --- & 3.3 & 19 & 1.2 & --- & 5.2 & 23 & 1.2 & --- \\
$\ce{O(^1D)}/\ce{O(^5S)}$ & $(6300+6364)/1356$ & 7.1 & 9.2 & 37 & --- & 7.3 & 6.3 & 29 & --- & 7.0 & 7.8 & 32 & --- \\
$\ce{O(^1D)}/\ce{O(^3S)}$ & $(6300+6364)/1304$ & 0.89 & 20 & 9.3 & --- & 0.16 & 14 & 7.2 & --- & 0.5 & 17 & 8 & --- \\
$\ce{O(^1D)}/\ce{O({3p}^5P)}$ & $(6300+6364)/7774$ & 16 & 17 & 15 & --- & 15 & 9.6 & 18 & --- & 15 & 13 & 15 & --- \\
$\ce{O(^1D)}/\ce{O({3p}^3P)}$ & $(6300+6364)/8446$ & 3.3 & 33 & 7.6 & --- & 1.1 & 20 & 6.9 & --- & 2 & 27 & 7 & --- \\
$\ce{O(^5S)}/\ce{O(^3S)}$ & $1356/1304$ & 0.11 & 2.2 & 0.25 & 0.91 & 0.022 & 2.2 & 0.25 & 1.0 & 0.07 & 2.2 & 0.25 & 0.97 \\
$\ce{O(^5S)}/\ce{O(^1S)}$ & $1356/5577$ & 0.94 & 2.9 & 0.033 & 0.026 & 0.046 & 2.9 & 0.043 & 0.046 & 0.74 & 3.0 & 0.038 & 0.034
\enddata
\end{deluxetable}

\begin{figure}
    \centering
    \includegraphics[width=0.6\textwidth]{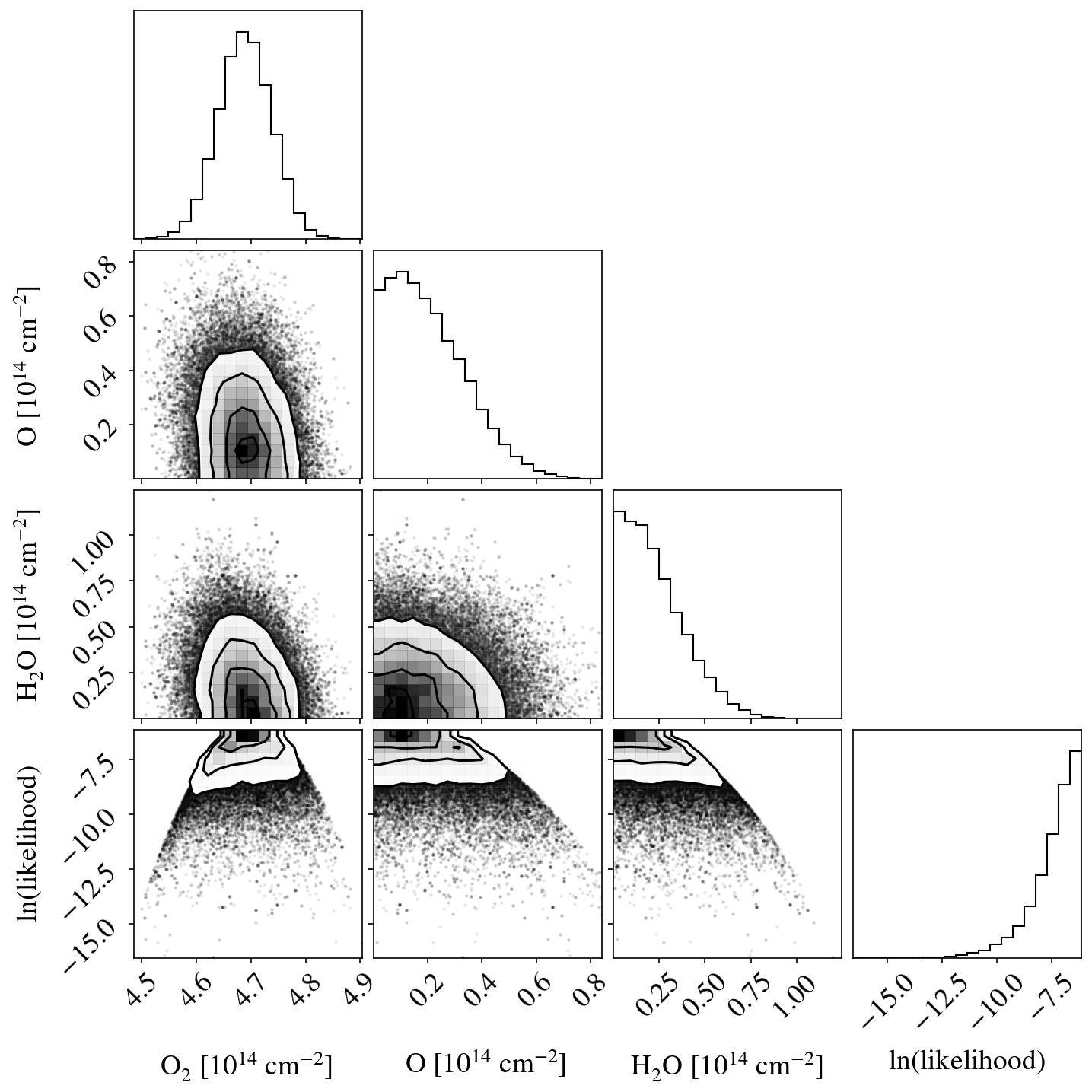}
    \caption{Example MCMC output showing the individual and joint posterior distributions for the column densities of the three atmospheric species, based on the measurements and upper limits for the Ganymede average case. In the fits, the electron energies and densities are fixed at the values described in the text, and the column densities of the three species are the free parameters in the fit.}
    \label{fig:mcmc_example}
\end{figure}

\section{Best-fit model properties}

\setlength{\tabcolsep}{2pt}
\begin{deluxetable}{@{\extracolsep{4pt}}lcccccccccccc@{}}
\caption{Best-fit models for Europa, Ganymede, and Callisto.\textsuperscript{a} \label{tbl:CombinedModel}}
\tabletypesize{\footnotesize}
\centering
\tablehead{
\colhead{} & \multicolumn{3}{c}{Column Density} & \multicolumn{8}{c}{Aurora Surface Brightness} \\
\cline{2-4} \cline{5-12}
\colhead{} & \colhead{\ce{O2}} & \colhead{\ce{O}} & \colhead{\ce{H2O}} & \colhead{$6300+6364\,\si{\angstrom}$} & \colhead{\SI{5577}{\angstrom}} & \colhead{\SI{7774}{\angstrom}} & \colhead{\SI{8446}{\angstrom}} & \colhead{\SI{6563}{\angstrom}} & \colhead{\SI{1304}{\angstrom}} & \colhead{\SI{1356}{\angstrom}} & \colhead{\SI{1216}{\angstrom}} \\[-3pt]
 Date & \multicolumn{3}{c}{[$\times\, 10^{14}\,\si{cm^{-2}}$]} & \ce{O(^1D)} & \ce{O(^1S)} & \ce{O({3p}^5P)} & \ce{O({3p}^3P)} & $\text{H-}\mathrm{\alpha}$ & \ce{O(^3S)} & \ce{O(^5S)} & $\text{Ly-}\mathrm{\alpha}$
}
\startdata
\textbf{Europa}\\
\quad 2021-05-20 &\\ 
\quad\quad \textit{Result w/Uncertainties} & \num{4.0 \pm 0.1} & $< 0.4$ & $< 0.9$ & --- & --- & --- & --- & --- & --- & --- & ---\\
\quad\quad \textit{Max-Likelihood Model} & 4.0 & 0.0003 & 0.73 & 763 & 31 & 44 & 24 & 6 & 38 & 83 & 11 \\
\quad\quad \textit{Measured Brightnesses} & --- & --- & --- & \num{763 \pm 16} & \num{37 \pm 4} & \num{30 \pm 15} & $< 22$ & $< 12$ & --- & --- & --- \\
\quad 2021-06-21 & \\
\quad\quad \textit{Result w/Uncertainties} & \num{3.7 \pm 0.2} & \num{1.8 \pm 0.8} & $< 0.8$ \\
\quad\quad \textit{Max-Likelihood Model} & 3.7 & 1.8 & 0.16 & 788 & 40 & 46 & 49 & 1.2 & 137 & 89 & 2.3 \\
\quad\quad \textit{Measured Brightnesses} & --- & --- & --- & \num{789 \pm 13} & \num{42 \pm 5} & $< 40$ & $< 80$ & $< 12$ & --- & --- & --- \\
\quad 2021-07-16\\
\quad\quad \textit{Result w/Uncertainties} & \num{3.7 \pm 0.2} & $< 1.2$ & $< 1.1$ & --- & --- & --- & --- & --- & --- & --- & ---\\
\quad\quad \textit{Max-Likelihood Model} & 3.7 & 0.37 & 0.35 & 717 & 30 & 42 & 27 & 3 & 56 & 79 & 5 \\
\quad\quad \textit{Measured Brightnesses} & --- & --- & --- & \num{717 \pm 16} & \num{35 \pm 8} & $< 60$ & $< 60$ & $< 18$ & --- & --- & --- \\
\quad Average\textsuperscript{b}\\
\quad\quad \textit{Result w/Uncertainties} & \num{4.1 \pm 0.1} & $< 0.1$ & \num{1.2 \pm 0.5} \\
\quad\quad \textit{Max-Likelihood Model} & 4.1 & 0.02 & 1.2 & 777 & 34 & 45 & 25 & 10 & 40 & 84 & 18 \\
\quad\quad \textit{Measured Brightnesses} & --- & --- & --- & \num{776 \pm 10} & \num{40 \pm 3} & \num{30 \pm 15} & $< 22$ & $< 12$ & \num{40 \pm 4} & \num{80 \pm 8} & --- \\[3pt]
\textbf{Ganymede}\\
\quad 1998-11-15 \\
\quad\quad \textit{Result w/Uncertainties} & \num{4.5 \pm 0.1} & $< 3$ & $< 0.6$ & --- & --- & --- & --- & --- & --- & --- & ---\\
\quad\quad \textit{Max-Likelihood Model} & 4.5 & 0.001 & 0.004 & 182 & 10 & 19 & 9 & 0 & 13 & 29 & 0 \\
\quad\quad \textit{Measured Brightnesses} & --- & --- & --- & \num{181.9 \pm 3.4} & \num{9.2 \pm 1.6} & --- & --- & $< 6$ & --- & --- & --- \\
\quad 2018-06-15 \\
\quad\quad \textit{Result w/Uncertainties} & \num{3.2 \pm 0.3} & $< 8$ & \num{16 \pm 3} & --- & --- & --- & --- & --- & --- & --- & ---\\
\quad\quad \textit{Max-Likelihood Model} & 3.2 & 2.7 & 16 & 164 & 33 & 16 & 15 & 50 & 40 & 22 & 108 \\
\quad\quad \textit{Measured Brightnesses} & --- & --- & --- & \num{163 \pm 9} & \num{41 \pm 4} & $< 26$ & $< 20$ & $< 30$ & --- & --- & --- \\
\quad 2021-06-08 \\
\quad\quad \textit{Result w/Uncertainties} & \num{4.8 \pm 0.1} & $< 0.8$ & $< 0.5$ & --- & --- & --- & --- & --- & --- & --- & ---\\
\quad\quad \textit{Max-Likelihood Model} & 4.8 & 0.005 & 0.45 & 193 & 11 & 20 & 10 & 1 & 14 & 31 & 3 \\
\quad\quad \textit{Measured Brightnesses} & --- & --- & --- & \num{192.6 \pm 1.8} & \num{12.5 \pm 0.6} & \num{20 \pm 2} & \num{7.3 \pm 1.4} & $< 1.8$ & --- & --- & --- \\
\quad Average\textsuperscript{b} \\
\quad\quad \textit{Result w/Uncertainties} & \num{4.7 \pm 0.1} & $< 0.3$ & $< 0.3$ & --- & --- & --- & --- & --- & --- & --- & ---\\
\quad\quad \textit{Max-Likelihood Model} & 4.7 & 0.12 & 0.04 & 188 & 10 & 20 & 10 & 0.1 & 15 & 30 & 0.3 \\
\quad\quad \textit{Measured Brightnesses} & --- & --- & --- & \num{187 \pm 2} & \num{11 \pm 1} & \num{20 \pm 2} & \num{7.3 \pm 1.4} & $< 2$ & \num{15 \pm 2} & \num{36 \pm 2} & --- \\[3pt]
\textbf{Callisto}\\
\quad 2021-07-04 &\\
\quad\quad \textit{Result w/Uncertainties} & \num{40 \pm 9} & --- & --- & 9.3 & 0.4 & 0.7 & 0.3 & 0 & 0.5 & 1.2 & 0.3 \\
\quad\quad \textit{Measured Brightnesses} & --- & --- & --- & \num{9.4 \pm 2.1} & $< 2$ & $< 8$ & $< 4$ & $< 3$ & --- & --- & --- \\[3pt]
\enddata
\vspace*{3pt}
\footnotesize{\textsuperscript{a}Our reported best-fit model with uncertainties is given under the heading ``Result w/Uncertainties''. However, because the column densities are often upper limits, we also reported the single model that maximizes the likelihood function along with the auroral surface brightnesses of that model.}\\
\footnotesize{\textsuperscript{b}Average auroral brightness are averaged over values from this work and from the literature \citep{dKB2018, dKB2019, roth2016}.}
\end{deluxetable}

\clearpage
\bibliography{IcySatAurora}{}

\begin{thebibliography}{}
\expandafter\ifx\csname natexlab\endcsname\relax\def\natexlab#1{#1}\fi
\providecommand{\url}[1]{\href{#1}{#1}}
\providecommand{\dodoi}[1]{doi:~\href{http://doi.org/#1}{\nolinkurl{#1}}}
\providecommand{\doeprint}[1]{\href{http://ascl.net/#1}{\nolinkurl{http://ascl.net/#1}}}
\providecommand{\doarXiv}[1]{\href{https://arxiv.org/abs/#1}{\nolinkurl{https://arxiv.org/abs/#1}}}

\bibitem[{{Ajello}(1971)}]{ajello1971}
{Ajello}, J.~M. 1971, \jcp, 55, 3169, \dodoi{10.1063/1.1676564}

\bibitem[{{Alday} {et~al.}(2017){Alday}, {Roth}, {Ivchenko}, {Retherford},
  {Becker}, {Molyneux}, \& {Saur}}]{alday2017}
{Alday}, J., {Roth}, L., {Ivchenko}, N., {et~al.} 2017, \planss, 148, 35,
  \dodoi{10.1016/j.pss.2017.10.006}

\bibitem[{{Bagenal} \& {Delamere}(2011)}]{bagenal2011}
{Bagenal}, F., \& {Delamere}, P.~A. 2011, Journal of Geophysical Research
  (Space Physics), 116, A05209, \dodoi{10.1029/2010JA016294}

\bibitem[{{Bagenal} \& {Dols}(2020)}]{bagenal2020}
{Bagenal}, F., \& {Dols}, V. 2020, Journal of Geophysical Research (Space
  Physics), 125, e27485, \dodoi{10.1029/2019JA027485}

\bibitem[{{Bagenal} {et~al.}(2015){Bagenal}, {Sidrow}, {Wilson}, {Cassidy},
  {Dols}, {Crary}, {Steffl}, {Delamere}, {Kurth}, \& {Paterson}}]{bagenal2015}
{Bagenal}, F., {Sidrow}, E., {Wilson}, R.~J., {et~al.} 2015, \icarus, 261, 1,
  \dodoi{10.1016/j.icarus.2015.07.036}

\bibitem[{{Beenakker} {et~al.}(1974){Beenakker}, {Heer}, {Krop}, \&
  {M{\"o}hlmann}}]{beenakker1974}
{Beenakker}, C.~I.~M., {Heer}, F.~J.~D., {Krop}, H.~B., \& {M{\"o}hlmann},
  G.~R. 1974, Chemical Physics, 6, 445, \dodoi{10.1016/0301-0104(74)85028-7}

\bibitem[{{Belcher}(1983)}]{belcher1983}
{Belcher}, J.~W. 1983, in Physics of the Jovian Magnetosphere (Cambridge
  University Press), 68--105

\bibitem[{{Belton} {et~al.}(1996){Belton}, {Head}, {Ingersoll}, {Greeley},
  {McEwen}, {Klaasen}, {Senske}, {Pappalardo}, {Collins}, {Vasavada},
  {Sullivan}, {Simonelli}, {Geissler}, {Carr}, {Davies}, {Veverka}, {Gierasch},
  {Banfield}, {Bell}, {Chapman}, {Anger}, {Greenberg}, {Neukum}, {Pilcher},
  {Beebe}, {Burns}, {Fanale}, {Ip}, {Johnson}, {Morrison}, {Moore}, {Orton},
  {Thomas}, \& {West}}]{belton1996}
{Belton}, M.~J.~S., {Head}, J.~W., I., {Ingersoll}, A.~P., {et~al.} 1996,
  Science, 274, 377, \dodoi{10.1126/science.274.5286.377}

\bibitem[{{Borucki} {et~al.}(1996){Borucki}, {McKay}, {Jebens}, {Lakkaraju}, \&
  {Vanajakshi}}]{borucki1996}
{Borucki}, W.~J., {McKay}, C.~P., {Jebens}, D., {Lakkaraju}, H.~S., \&
  {Vanajakshi}, C.~T. 1996, \icarus, 123, 336, \dodoi{10.1006/icar.1996.0162}

\bibitem[{{Bougher} {et~al.}(2017){Bougher}, {Brain}, {Fox}, {Francisco},
  {Simon-Wedlund}, \& {Withers}}]{bougher2017}
{Bougher}, S.~W., {Brain}, D.~A., {Fox}, J.~L., {et~al.} 2017, in The
  Atmosphere and Climate of Mars, ed. R.~M. {Haberle}, R.~T. {Clancy},
  F.~{Forget}, M.~D. {Smith}, \& R.~W. {Zurek} (Cambridge University Press),
  405--432, \dodoi{10.1017/9781139060172.014}

\bibitem[{{Broadfoot} {et~al.}(1979){Broadfoot}, {Belton}, {Takacs}, {Sandel},
  {Shemansky}, {Holberg}, {Ajello}, {Atreya}, {Donahue}, {Moos}, {Bertaux},
  {Blamont}, {Strobel}, {McConnell}, {Dalgarno}, {Goody}, \&
  {McElroy}}]{broadfoot1979}
{Broadfoot}, A.~L., {Belton}, M.~J.~S., {Takacs}, P.~Z., {et~al.} 1979,
  Science, 204, 979, \dodoi{10.1126/science.204.4396.979}

\bibitem[{{Brown} {et~al.}(1980){Brown}, {Augustyniak}, {Lanzerotti},
  {Johnson}, \& {Evatt}}]{brown1980}
{Brown}, W.~L., {Augustyniak}, W.~M., {Lanzerotti}, L.~J., {Johnson}, R.~E., \&
  {Evatt}, R. 1980, \prl, 45, 1632, \dodoi{10.1103/PhysRevLett.45.1632}

\bibitem[{{Buton} {et~al.}(2013){Buton}, {Copin}, {Aldering}, {Antilogus},
  {Aragon}, {Bailey}, {Baltay}, {Bongard}, {Canto}, {Cellier-Holzem},
  {Childress}, {Chotard}, {Fakhouri}, {Gangler}, {Guy}, {Hsiao}, {Kerschhaggl},
  {Kowalski}, {Loken}, {Nugent}, {Paech}, {Pain}, {P{\'e}contal}, {Pereira},
  {Perlmutter}, {Rabinowitz}, {Rigault}, {Runge}, {Scalzo}, {Smadja}, {Tao},
  {Thomas}, {Weaver}, {Wu}, \& {Nearby SuperNova Factory}}]{Buton2013}
{Buton}, C., {Copin}, Y., {Aldering}, G., {et~al.} 2013, Astronomy and
  Astrophysics, 549, \dodoi{10.1051/0004-6361/201219834}

\bibitem[{{Carberry Mogan} {et~al.}(2021){Carberry Mogan}, {Tucker}, {Johnson},
  {Vorburger}, {Galli}, {Marchand}, {Tafuni}, {Kumar}, {Sahin}, \&
  {Sreenivasan}}]{carberry-mogan2021}
{Carberry Mogan}, S.~R., {Tucker}, O.~J., {Johnson}, R.~E., {et~al.} 2021,
  \icarus, 368, 114597, \dodoi{10.1016/j.icarus.2021.114597}

\bibitem[{{Carlson}(1999)}]{carlson1999}
{Carlson}, R.~W. 1999, Science, 283, 820, \dodoi{10.1126/science.283.5403.820}

\bibitem[{{Carlson} {et~al.}(1973){Carlson}, {Bhattacharyya}, {Smith},
  {Johnson}, {Hidayat}, {Smith}, {Taylor}, {O'Leary}, \&
  {Brinkmann}}]{carlson1973}
{Carlson}, R.~W., {Bhattacharyya}, J.~C., {Smith}, B.~A., {et~al.} 1973,
  Science, 182, 53, \dodoi{10.1126/science.182.4107.53}

\bibitem[{Cassidy {et~al.}(2008)Cassidy, Johnson, Geissler, \&
  Leblanc}]{Cassidy2008}
Cassidy, T.~A., Johnson, R.~E., Geissler, P.~E., \& Leblanc, F. 2008, Journal
  of Geophysical Research: Planets, 113,
  \dodoi{https://doi.org/10.1029/2007JE002955}

\bibitem[{{Cassidy} {et~al.}(2013){Cassidy}, {Paranicas}, {Shirley}, {Dalton},
  {Teolis}, {Johnson}, {Kamp}, \& {Hendrix}}]{cassidy2013}
{Cassidy}, T.~A., {Paranicas}, C.~P., {Shirley}, J.~H., {et~al.} 2013, \planss,
  77, 64, \dodoi{10.1016/j.pss.2012.07.008}

\bibitem[{{Cunningham} {et~al.}(2015){Cunningham}, {Spencer}, {Feldman},
  {Strobel}, {France}, \& {Osterman}}]{cunningham2015}
{Cunningham}, N.~J., {Spencer}, J.~R., {Feldman}, P.~D., {et~al.} 2015,
  \icarus, 254, 178, \dodoi{10.1016/j.icarus.2015.03.021}

\bibitem[{{de Kleer} \& {Brown}(2018)}]{dKB2018}
{de Kleer}, K., \& {Brown}, M.~E. 2018, \aj, 156, 167,
  \dodoi{10.3847/1538-3881/aadae8}

\bibitem[{{de Kleer} \& {Brown}(2019)}]{dKB2019}
---. 2019, Research Notes of the American Astronomical Society, 3, 27,
  \dodoi{10.3847/2515-5172/ab0289}

\bibitem[{{Eviatar} {et~al.}(2001){Eviatar}, {Strobel}, {Wolven}, {Feldman},
  {McGrath}, \& {Williams}}]{eviatar2001_aurora}
{Eviatar}, A., {Strobel}, D.~F., {Wolven}, B.~C., {et~al.} 2001, \apj, 555,
  1013, \dodoi{10.1086/321510}

\bibitem[{{Feistel} \& {Wagner}(2007)}]{feistel2007}
{Feistel}, R., \& {Wagner}, W. 2007, \gca, 71, 36,
  \dodoi{10.1016/j.gca.2006.08.034}

\bibitem[{{Feldman} {et~al.}(2000){Feldman}, {McGrath}, {Strobel}, {Moos},
  {Retherford}, \& {Wolven}}]{feldman2000}
{Feldman}, P.~D., {McGrath}, M.~A., {Strobel}, D.~F., {et~al.} 2000, \apj, 535,
  1085, \dodoi{10.1086/308889}

\bibitem[{{Foreman-Mackey} {et~al.}(2013){Foreman-Mackey}, {Hogg}, {Lang}, \&
  {Goodman}}]{emcee}
{Foreman-Mackey}, D., {Hogg}, D.~W., {Lang}, D., \& {Goodman}, J. 2013, \pasp,
  125, 306, \dodoi{10.1086/670067}

\bibitem[{{Goodman} \& {Weare}(2010)}]{goodman2010}
{Goodman}, J., \& {Weare}, J. 2010, Communications in Applied Mathematics and
  Computational Science, 5, 65, \dodoi{10.2140/camcos.2010.5.65}

\bibitem[{{Gronoff} {et~al.}(2008){Gronoff}, {Lilensten}, {Simon},
  {Barth{\'e}lemy}, {Leblanc}, \& {Dutuit}}]{gronoff2008}
{Gronoff}, G., {Lilensten}, J., {Simon}, C., {et~al.} 2008, \aap, 482, 1015,
  \dodoi{10.1051/0004-6361:20077503}

\bibitem[{{Gulcicek} \& {Doering}(1987)}]{gulcicek1987}
{Gulcicek}, E.~E., \& {Doering}, J.~P. 1987, \jgr, 92, 3445,
  \dodoi{10.1029/JA092iA04p03445}

\bibitem[{{Gulcicek} {et~al.}(1988){Gulcicek}, {Doering}, \&
  {Vaughan}}]{gulcicek1988}
{Gulcicek}, E.~E., {Doering}, J.~P., \& {Vaughan}, S.~O. 1988, \jgr, 93, 5885,
  \dodoi{10.1029/JA093iA06p05885}

\bibitem[{{Gurnett} {et~al.}(2000){Gurnett}, {Persoon}, {Kurth}, {Roux}, \&
  {Bolton}}]{gurnett2000}
{Gurnett}, D.~A., {Persoon}, A.~M., {Kurth}, W.~S., {Roux}, A., \& {Bolton},
  S.~J. 2000, \grl, 27, 1867, \dodoi{10.1029/2000GL003751}

\bibitem[{{Hall} {et~al.}(1998){Hall}, {Feldman}, {McGrath}, \&
  {Strobel}}]{hall1998}
{Hall}, D.~T., {Feldman}, P.~D., {McGrath}, M.~A., \& {Strobel}, D.~F. 1998,
  \apj, 499, 475, \dodoi{10.1086/305604}

\bibitem[{{Hall} {et~al.}(1995){Hall}, {Strobel}, {Feldman}, {McGrath}, \&
  {Weaver}}]{hall1995}
{Hall}, D.~T., {Strobel}, D.~F., {Feldman}, P.~D., {McGrath}, M.~A., \&
  {Weaver}, H.~A. 1995, \nat, 373, 677, \dodoi{10.1038/373677a0}

\bibitem[{{Hansell} {et~al.}(1995){Hansell}, {Wells}, \&
  {Hunten}}]{Hansell1995}
{Hansell}, S.~A., {Wells}, W.~K., \& {Hunten}, D.~M. 1995, \icarus, 117, 345,
  \dodoi{10.1006/icar.1995.1160}

\bibitem[{{Hill} {et~al.}(1974){Hill}, {Dessler}, \& {Michel}}]{hill1974}
{Hill}, T.~W., {Dessler}, A.~J., \& {Michel}, F.~C. 1974, \grl, 1, 3,
  \dodoi{10.1029/GL001i001p00003}

\bibitem[{{Ip}(1996)}]{ip1996}
{Ip}, W.~H. 1996, \icarus, 120, 317, \dodoi{10.1006/icar.1996.0052}

\bibitem[{{Itikawa}(2002)}]{itikawa2002}
{Itikawa}, Y. 2002, Journal of Physical and Chemical Reference Data, 31, 749,
  \dodoi{10.1063/1.1481879}

\bibitem[{{Itikawa} \& {Mason}(2005)}]{itikawa2005}
{Itikawa}, Y., \& {Mason}, N. 2005, Journal of Physical and Chemical Reference
  Data, 34, 1, \dodoi{10.1063/1.1799251}

\bibitem[{{Johnson} {et~al.}(1982){Johnson}, {Lanzerotti}, \&
  {Brown}}]{johnson1982}
{Johnson}, R.~E., {Lanzerotti}, L.~J., \& {Brown}, W.~L. 1982, Nuclear
  Instruments and Methods in Physics Research, 198, 147,
  \dodoi{10.1016/0167-5087(82)90066-7}

\bibitem[{Johnson {et~al.}(1981)Johnson, Lanzerotti, Brown, \&
  Armstrong}]{johnson1981}
Johnson, R.~E., Lanzerotti, L.~J., Brown, W.~L., \& Armstrong, T.~P. 1981,
  Science, 212, 1027, \dodoi{10.1126/science.212.4498.1027}

\bibitem[{{Johnson} {et~al.}(2019){Johnson}, {Oza}, {Leblanc}, {Schmidt},
  {Nordheim}, \& {Cassidy}}]{johnson2019}
{Johnson}, R.~E., {Oza}, A.~V., {Leblanc}, F., {et~al.} 2019, \ssr, 215, 20,
  \dodoi{10.1007/s11214-019-0582-1}

\bibitem[{{Kivelson} {et~al.}(2004){Kivelson}, {Bagenal}, {Kurth}, {Neubauer},
  {Paranicas}, \& {Saur}}]{kivelson2004}
{Kivelson}, M.~G., {Bagenal}, F., {Kurth}, W.~S., {et~al.} 2004, in Jupiter.
  The Planet, Satellites and Magnetosphere, ed. F.~{Bagenal}, T.~E. {Dowling},
  \& W.~B. {McKinnon}, Vol.~1 (Cambridge University Press), 513--536

\bibitem[{{Kliore} {et~al.}(2002){Kliore}, {Anabtawi}, {Herrera}, {Asmar},
  {Nagy}, {Hinson}, \& {Flasar}}]{kliore2002}
{Kliore}, A.~J., {Anabtawi}, A., {Herrera}, R.~G., {et~al.} 2002, Journal of
  Geophysical Research (Space Physics), 107, 1407, \dodoi{10.1029/2002JA009365}

\bibitem[{Kurth {et~al.}(2022)Kurth, Sulaiman, Hospodarsky, B.~H.~Mauk, Clark,
  Allegrini, Valek, Connerney, Waite, Bolton, Imai, Santolik, Li, Duling, Saur,
  \& Louis}]{kurth2022}
Kurth, W.~S., Sulaiman, A.~H., Hospodarsky, G.~B., {et~al.} 2022, Geophysical
  Research Letters, n/a, e2022GL098591,
  \dodoi{https://doi.org/10.1029/2022GL098591}

\bibitem[{{Laher} \& {Gilmore}(1990)}]{laher1990}
{Laher}, R.~R., \& {Gilmore}, F.~R. 1990, Journal of Physical and Chemical
  Reference Data, 19, 277, \dodoi{10.1063/1.555872}

\bibitem[{{Lanzerotti} {et~al.}(1978){Lanzerotti}, {Brown}, {Poate}, \&
  {Augustyniak}}]{lanzerotti1978}
{Lanzerotti}, L.~J., {Brown}, W.~L., {Poate}, J.~M., \& {Augustyniak}, W.~M.
  1978, \grl, 5, 155, \dodoi{10.1029/GL005i002p00155}

\bibitem[{{Leblanc} {et~al.}(2017){Leblanc}, {Oza}, {Leclercq}, {Schmidt},
  {Cassidy}, {Modolo}, {Chaufray}, \& {Johnson}}]{leblanc2017}
{Leblanc}, F., {Oza}, A.~V., {Leclercq}, L., {et~al.} 2017, \icarus, 293, 185,
  \dodoi{10.1016/j.icarus.2017.04.025}

\bibitem[{{LeClair} \& {McConkey}(1994)}]{leclair1994}
{LeClair}, L.~R., \& {McConkey}, J.~W. 1994, Journal of Physics B Atomic
  Molecular Physics, 27, 4039, \dodoi{10.1088/0953-4075/27/17/026}

\bibitem[{{Liang} {et~al.}(2005){Liang}, {Lane}, {Pappalardo}, {Allen}, \&
  {Yung}}]{liang2005}
{Liang}, M.-C., {Lane}, B.~F., {Pappalardo}, R.~T., {Allen}, M., \& {Yung},
  Y.~L. 2005, Journal of Geophysical Research (Planets), 110, E02003,
  \dodoi{10.1029/2004JE002322}

\bibitem[{Lorenz {et~al.}(2019)Lorenz, Imai, Takahashi, Sato, Yamazaki, Sato,
  Imamura, Satoh, \& Nakamura}]{lorenz2019}
Lorenz, R.~D., Imai, M., Takahashi, Y., {et~al.} 2019, Geophysical Research
  Letters, 46, 7955, \dodoi{https://doi.org/10.1029/2019GL083311}

\bibitem[{{Marconi}(2007)}]{marconi2007}
{Marconi}, M.~L. 2007, \icarus, 190, 155, \dodoi{10.1016/j.icarus.2007.02.016}

\bibitem[{McCully {et~al.}(2018)McCully, Crawford, Kovacs, Tollerud, Betts,
  Bradley, Craig, Turner, Streicher, Sipocz, \& et~al.}]{mccully18}
McCully, C., Crawford, S., Kovacs, G., {et~al.} 2018, {Astro-SCRAPPY}.
\newblock \url{https://zenodo.org/record/1482019}

\bibitem[{{McGrath} {et~al.}(2013){McGrath}, {Jia}, {Retherford}, {Feldman},
  {Strobel}, \& {Saur}}]{mcgrath2013}
{McGrath}, M.~A., {Jia}, X., {Retherford}, K., {et~al.} 2013, Journal of
  Geophysical Research (Space Physics), 118, 2043, \dodoi{10.1002/jgra.50122}

\bibitem[{{Molyneux} {et~al.}(2018){Molyneux}, {Nichols}, {Bannister}, {Bunce},
  {Clarke}, {Cowley}, {G{\'e}rard}, {Grodent}, {Milan}, \&
  {Paty}}]{molyneux2018}
{Molyneux}, P.~M., {Nichols}, J.~D., {Bannister}, N.~P., {et~al.} 2018, Journal
  of Geophysical Research (Space Physics), 123, 3777,
  \dodoi{10.1029/2018JA025243}

\bibitem[{{Mumma} {et~al.}(1972){Mumma}, {Stone}, {Borst}, \&
  {Zipf}}]{mumma1972}
{Mumma}, M.~J., {Stone}, E.~J., {Borst}, W.~L., \& {Zipf}, E.~C. 1972, \jcp,
  57, 68, \dodoi{10.1063/1.1678019}

\bibitem[{{Oliversen} {et~al.}(2001){Oliversen}, {Scherb}, {Smyth}, {Freed},
  {Woodward}, {Marconi}, {Retherford}, {Lupie}, \&
  {Morgenthaler}}]{oliversen2001}
{Oliversen}, R.~J., {Scherb}, F., {Smyth}, W.~H., {et~al.} 2001, \jgr, 106,
  26183, \dodoi{10.1029/2000JA002507}

\bibitem[{{Oza} {et~al.}(2018){Oza}, {Johnson}, \& {Leblanc}}]{oza2018}
{Oza}, A.~V., {Johnson}, R.~E., \& {Leblanc}, F. 2018, \icarus, 305, 50,
  \dodoi{10.1016/j.icarus.2017.12.032}

\bibitem[{{Oza} {et~al.}(2019){Oza}, {Leblanc}, {Johnson}, {Schmidt},
  {Leclercq}, {Cassidy}, \& {Chaufray}}]{oza2019}
{Oza}, A.~V., {Leblanc}, F., {Johnson}, R.~E., {et~al.} 2019, \planss, 167, 23,
  \dodoi{10.1016/j.pss.2019.01.006}

\bibitem[{{Pavlov} \& {Berrington}(1999)}]{Pavlov1999}
{Pavlov}, A.~V., \& {Berrington}, K.~A. 1999, Annales Geophysicae, 17, 919,
  \dodoi{10.1007/s00585-999-0919-2}

\bibitem[{{Phipps} \& {Bagenal}(2021)}]{phipps2021}
{Phipps}, P., \& {Bagenal}, F. 2021, Journal of Geophysical Research (Space
  Physics), 126, e28713, \dodoi{10.1029/2020JA028713}

\bibitem[{{Plainaki} {et~al.}(2013){Plainaki}, {Milillo}, {Mura}, {Saur},
  {Orsini}, \& {Massetti}}]{plainaki2013}
{Plainaki}, C., {Milillo}, A., {Mura}, A., {et~al.} 2013, \planss, 88, 42,
  \dodoi{10.1016/j.pss.2013.08.011}

\bibitem[{{Roth}(2021)}]{roth2021_europa}
{Roth}, L. 2021, \grl, 48, e94289, \dodoi{10.1029/2021GL094289}

\bibitem[{{Roth} {et~al.}(2021){Roth}, {Ivchenko}, {Gladstone}, {Saur},
  {Grodent}, {Bonfond}, {Molyneux}, \& {Retherford}}]{roth2021_ganymede}
{Roth}, L., {Ivchenko}, N., {Gladstone}, G.~R., {et~al.} 2021, Nature
  Astronomy, 5, 1043, \dodoi{10.1038/s41550-021-01426-9}

\bibitem[{{Roth} {et~al.}(2014{\natexlab{a}}){Roth}, {Retherford}, {Saur},
  {Strobel}, {Feldman}, {McGrath}, \& {Nimmo}}]{roth2014_apocenter}
{Roth}, L., {Retherford}, K.~D., {Saur}, J., {et~al.} 2014{\natexlab{a}},
  Proceedings of the National Academy of Science, 111, E5123,
  \dodoi{10.1073/pnas.1416671111}

\bibitem[{{Roth} {et~al.}(2014{\natexlab{b}}){Roth}, {Saur}, {Retherford},
  {Feldman}, \& {Strobel}}]{roth2014}
{Roth}, L., {Saur}, J., {Retherford}, K.~D., {Feldman}, P.~D., \& {Strobel},
  D.~F. 2014{\natexlab{b}}, \icarus, 228, 386,
  \dodoi{10.1016/j.icarus.2013.10.009}

\bibitem[{{Roth} {et~al.}(2016){Roth}, {Saur}, {Retherford}, {Strobel},
  {Feldman}, {McGrath}, {Spencer}, {Bl{\"o}cker}, \& {Ivchenko}}]{roth2016}
{Roth}, L., {Saur}, J., {Retherford}, K.~D., {et~al.} 2016, Journal of
  Geophysical Research (Space Physics), 121, 2143, \dodoi{10.1002/2015JA022073}

\bibitem[{{Saur} {et~al.}(1998){Saur}, {Strobel}, \& {Neubauer}}]{saur1998}
{Saur}, J., {Strobel}, D.~F., \& {Neubauer}, F.~M. 1998, \jgr, 103, 19947,
  \dodoi{10.1029/97JE03556}

\bibitem[{{Saur} {et~al.}(2015){Saur}, {Duling}, {Roth}, {Jia}, {Strobel},
  {Feldman}, {Christensen}, {Retherford}, {McGrath}, {Musacchio}, {Wennmacher},
  {Neubauer}, {Simon}, \& {Hartkorn}}]{saur2015}
{Saur}, J., {Duling}, S., {Roth}, L., {et~al.} 2015, Journal of Geophysical
  Research (Space Physics), 120, 1715, \dodoi{10.1002/2014JA020778}

\bibitem[{{Schmidt} {et~al.}(in press){Schmidt}, {Sharov}, {de Kleer},
  {Schneider}, {de Pater}, {Phipps}, {Conrad}, {Moore}, {Withers}, {Spencer},
  {Morgenthaler}, {Ilyin}, {Strassmeier}, {Veillet}, {Hill}, \&
  {Brown}}]{schmidt2022}
{Schmidt}, C., {Sharov}, M., {de Kleer}, K., {et~al.} in press, \psj

\bibitem[{{Schulman} {et~al.}(1985){Schulman}, {Sharpton}, {Chung}, {Lin}, \&
  {Anderson}}]{schulman1985}
{Schulman}, M.~B., {Sharpton}, F.~A., {Chung}, S., {Lin}, C.~C., \& {Anderson},
  L.~W. 1985, \pra, 32, 2100, \dodoi{10.1103/PhysRevA.32.2100}

\bibitem[{Shemansky {et~al.}(2014)Shemansky, Yung, Liu, Yoshii, Hansen,
  Hendrix, \& Esposito}]{shemansky2014}
Shemansky, D.~E., Yung, Y.~L., Liu, X., {et~al.} 2014, The Astrophysical
  Journal, 797, 84, \dodoi{10.1088/0004-637x/797/2/84}

\bibitem[{{Shematovich} {et~al.}(2005){Shematovich}, {Johnson}, {Cooper}, \&
  {Wong}}]{shematovich2005}
{Shematovich}, V.~I., {Johnson}, R.~E., {Cooper}, J.~F., \& {Wong}, M.~C. 2005,
  \icarus, 173, 480, \dodoi{10.1016/j.icarus.2004.08.013}

\bibitem[{{Sittler} \& {Strobel}(1987)}]{sittler1987}
{Sittler}, E.~C., \& {Strobel}, D.~F. 1987, \jgr, 92, 5741,
  \dodoi{10.1029/JA092iA06p05741}

\bibitem[{{Slanger} {et~al.}(2004){Slanger}, {Cosby}, {Huestis}, \&
  {Meier}}]{slanger2004}
{Slanger}, T.~G., {Cosby}, P.~C., {Huestis}, D.~L., \& {Meier}, R.~R. 2004,
  Journal of Geophysical Research (Space Physics), 109, A10309,
  \dodoi{10.1029/2004JA010556}

\bibitem[{{Smyth} \& {Marconi}(2006)}]{smyth2006}
{Smyth}, W.~H., \& {Marconi}, M.~L. 2006, \icarus, 181, 510,
  \dodoi{10.1016/j.icarus.2005.10.019}

\bibitem[{Streit {et~al.}(1976)Streit, Howard, Schmeltekopf, Davidson, \&
  Schiff}]{Streit1976}
Streit, G.~E., Howard, C.~J., Schmeltekopf, A.~L., Davidson, J.~A., \& Schiff,
  H.~I. 1976, The Journal of Chemical Physics, 65, 4761,
  \dodoi{10.1063/1.432930}

\bibitem[{{Strobel} {et~al.}(2002){Strobel}, {Saur}, {Feldman}, \&
  {McGrath}}]{strobel2002}
{Strobel}, D.~F., {Saur}, J., {Feldman}, P.~D., \& {McGrath}, M.~A. 2002,
  \apjl, 581, L51, \dodoi{10.1086/345803}

\bibitem[{{Teolis} {et~al.}(2017){Teolis}, {Plainaki}, {Cassidy}, \&
  {Raut}}]{teolis2017}
{Teolis}, B.~D., {Plainaki}, C., {Cassidy}, T.~A., \& {Raut}, U. 2017, Journal
  of Geophysical Research (Planets), 122, 1996, \dodoi{10.1002/2017JE005285}

\bibitem[{{van Dokkum}(2001)}]{vanDokkum2001}
{van Dokkum}, P.~G. 2001, Publications of the Astronomical Society of the
  Pacific, 113, 1420, \dodoi{10.1086/323894}

\bibitem[{{Vogt} {et~al.}(1994){Vogt}, {Allen}, {Bigelow}, {Bresee}, {Brown},
  {Cantrall}, {Conrad}, {Couture}, {Delaney}, {Epps}, {Hilyard}, {Hilyard},
  {Horn}, {Jern}, {Kanto}, {Keane}, {Kibrick}, {Lewis}, {Osborne},
  {Pardeilhan}, {Pfister}, {Ricketts}, {Robinson}, {Stover}, {Tucker}, {Ward},
  \& {Wei}}]{vogt1994}
{Vogt}, S.~S., {Allen}, S.~L., {Bigelow}, B.~C., {et~al.} 1994, in Society of
  Photo-Optical Instrumentation Engineers (SPIE) Conference Series, Vol. 2198,
  Instrumentation in Astronomy VIII, ed. D.~L. {Crawford} \& E.~R. {Craine},
  362, \dodoi{10.1117/12.176725}

\bibitem[{{Vorburger} {et~al.}(2022){Vorburger}, {Fatemi}, {Galli}, {Liuzzo},
  {Poppe}, \& {Wurz}}]{vorburger2022}
{Vorburger}, A., {Fatemi}, S., {Galli}, A., {et~al.} 2022, \icarus, 375,
  114810, \dodoi{10.1016/j.icarus.2021.114810}

\bibitem[{{Vorburger} \& {Wurz}(2021)}]{vorburger2021}
{Vorburger}, A., \& {Wurz}, P. 2021, Journal of Geophysical Research (Space
  Physics), 126, e29690, \dodoi{10.1029/2021JA029690}

\bibitem[{Wiese {et~al.}(1996)Wiese, Fuhr, \& Deters}]{wiese1996}
Wiese, W.~L., Fuhr, J.~R., \& Deters, T.~M. 1996, {Journal of Physical and
  Chemical Reference Data, Monograph No.~7}

\bibitem[{{Wolff} \& {Mendis}(1983)}]{wolff1983}
{Wolff}, R.~S., \& {Mendis}, D.~A. 1983, \jgr, 88, 4749,
  \dodoi{10.1029/JA088iA06p04749}

\bibitem[{{Woodman} {et~al.}(1979){Woodman}, {Cochran}, \&
  {Slavsky}}]{Woodman1979}
{Woodman}, J.~H., {Cochran}, W.~D., \& {Slavsky}, D.~B. 1979, Icarus, 37, 73,
  \dodoi{10.1016/0019-1035(79)90116-7}

\bibitem[{{Yung} \& {McElroy}(1977)}]{yung1977}
{Yung}, Y.~L., \& {McElroy}, M.~B. 1977, \icarus, 30, 97,
  \dodoi{10.1016/0019-1035(77)90124-5}

\end{thebibliography}
\bibliographystyle{aasjournal}

\end{document}